\documentclass[11pt]{article}
\usepackage[letterpaper,margin=1in]{geometry}
\usepackage[T1]{fontenc}
\usepackage{amsmath,amssymb,graphicx,booktabs,multirow}
\usepackage[authoryear,round]{natbib}
\usepackage[hidelinks,bookmarks=false]{hyperref}
\usepackage{cleveref}
\let\textAA\AA
\renewcommand{\AA}{\ifmmode\text{\textAA}\else\textAA\fi}
\DeclareUnicodeCharacter{00C5}{\AA}

\newcommand{\conv}{\ast}
\newcommand{\vq}{\mathbf{q}}
\newcommand{\vQ}{\mathbf{Q}}
\newcommand{\vR}{\mathbf{R}}

\newcommand{\vv}{\mathbf{v}}
\newcommand{\vx}{\mathbf{x}}
\newcommand{\vk}{\mathbf{k}}
\newcommand{\vxi}{\boldsymbol{\xi}}
\newcommand{\vchi}{\boldsymbol{\chi}}

\title{Dopant depth determination from momentum-resolved core-loss 4D-STEM data}
\author{\href{https://orcid.org/0000-0001-9296-5958}{Michael Deimetry}\thanks{Corresponding author.
\href{mailto:Michael.Deimetry@monash.edu}{Michael.Deimetry@monash.edu}
\newline
\textbf{Conflict of interest:} The author declares none.}\\[0.5em]
\small School of Physics and Astronomy\\
\small Monash University, Wellington Rd, Clayton, Victoria, 3800, Australia}
\date{}

\begin{document}
\maketitle
\begin{abstract}
Moving beyond specimen-projected micrographs in scanning transmission electron microscopy to determine a three-dimensional distribution of dopants is of vital importance for materials characterization and development. Within the Born approximation, Fourier transformation over probe position of a momentum-resolved core-loss 4D-STEM dataset reveals fringes whose frequency encodes the dopant's distance from the focal plane. We address two principal challenges which arise under experimental circumstances. First, fringe contrast is degraded by dynamical scattering, but can be partially recovered by acquiring off major zone axes. Second, the excessive doses typical of core-loss scattering can be alleviated using a Bayesian framework which, on a simulated Fe-doped strontium titanate, achieves nanometre to sub-nanometre depth precision from 100 to 1000 core-loss counts, depending on the signal-to-background.
\end{abstract}

\section{Introduction}

Scanning transmission electron microscopy (STEM) has become an indispensable tool for probing the structure and composition of materials at the atomic scale. Advances in aberration correction have made atomic-resolution imaging routine, while the advent of fast, highly sensitive pixelated detectors has enabled the recording of complete diffraction patterns at successive positions of a convergent electron probe  \citep{tate2016high,ercius20244d,philipp2022very}. The resulting four-dimensional STEM (4D-STEM) dataset resolves two real-space probe coordinates, $\vR$, and two reciprocal-space coordinates, $\vq$, in the objective lens back-focal plane. Coupled with increasingly accessible computational resources, this wealth of momentum-resolved information has given rise to a rapidly expanding repertoire of imaging and reconstruction techniques.

Despite these advances, determining the position of an individual dopant atom along the optic axis remains considerably more difficult than laterally. Conventional STEM images are, to leading order, projections of the specimen, leaving the majority of depth information only indirectly accessible through the diffraction of the electron wave. Consequently, interpreting the momentum distribution has  motivated a diverse range of approaches, each exploiting some property of electron scattering to recover depth information.

Among the most established is electron tomography, which reconstructs three-dimensional structure from projections acquired at different specimen orientations  \citep{weyland2004electron}. Spatial resolution is typically on the nanometre scale, although sub-nanometre resolution has also been demonstrated  \citep{scott2012electron}.  However, the experimentally restricted angular range introduces missing-wedge artefacts \citep{arslan2006reducing}, which can be partially alleviated by incorporating additional structural constraints \citep{leary2013compressed,saghi2011three}. Tomographic acquisition may also necessitate specialized specimen holders.

Alternatively, phase retrieval from 4D-STEM data acquired with substantial probe overlap, dubbed ptychography, comprises non-dynamical approaches such as Wigner distribution  deconvolution (WDD) \citep{rodenburg1992theory}, single-sideband \citep{pennycook2015efficient}, and the extended
ptychographical iterative engine (ePIE) \citep{maiden2009improved,rodenburg2004phase}, as well as those which explicitly account for dynamical diffraction, such as multislice ptychography \citep{maiden2012ptychographic,gao2017electron} and $S$-matrix reconstruction \citep{brown2018structure,terzoudis2023resolution,pelz2021phase}. 
By exploiting the momentum-resolved information available in 4D-STEM, these techniques offer a powerful route to recovering the specimen potential. 
Accounting for dynamical scattering introduces greater computational complexity compared with non-dynamical methods which are comparatively straightforward, but sacrifice fidelity.

When momentum resolution is unavailable, the finite depth of field of a convergent electron probe can be exploited in a through-focal acquisition. Depth sectioning, and similarly confocal imaging, exploit the axial localization of the probe intensity to interrogate successive planes within the specimen. In high-angle annular dark-field (HAADF) depth sectioning, a heavy dopant embedded in a lighter matrix produces an enhanced signal as the probe waist illuminates the heavy dopant \citep{borisevich2006depth}, permitting nanometre to sub-nanometre depth precision for isolated heavy atoms \citep{van2006three}. Core-loss depth sectioning, proposed by \cite{d2007depth}, removes this dependence on atomic weight by distinguishing dopants through their characteristic electronic excitation energies.

More recently, defocused STEM-EELS \citep{deimetry2026threedimensionaldopantdistributionsingle} has offered a means of recovering dopant depth without acquiring a through-focal series. Rather than tracking the variation in signal with focal position, the technique exploits the full geometry of the defocused Airy probe, rather than just the beam waist, to infer the dopant's axial location from a single acquisition. Like conventional depth sectioning, it requires no momentum resolution and has been demonstrated to achieve nanometre-scale depth precision.

The present work explores how this depth sensitivity can be enhanced by retaining the momentum information ordinarily discarded in STEM-EELS. Momentum-resolved measurements have long been recognized as a rich source of structural information, both in elastic scattering \citep{ophus2019four} and in the core-loss regime \citep{midgley1995energy,muller2017measurement}. 
Here, we demonstrate that the core-loss momentum distribution of a single atom encodes its three-dimensional position. 

We first consider an isolated atom in vacuum, for which the scattering process admits an analytical treatment within the Born approximation. 
Through straightforward manipulation of the dataset, the dopant's placement along the optic axis can be readily inferred, without iterative schemes. 
We then examine the extent to which this simple correspondence survives in crystalline specimens, where dynamical scattering complicates the interpretation. Strong on-zone-axis channelling is found to compromise depth retrieval which can be considerably mitigated by tilting the specimen away from major crystallographic axes.

The experimental feasibility of the proposed technique is ultimately limited by the scarcity of single-atom core-loss events. Acquiring fully momentum-resolved core-loss 4D-STEM datasets therefore presents a considerable challenge, particularly for radiation-sensitive specimens. Although single-atom core-loss detection has been demonstrated previously, such measurements have generally required carefully controlled conditions and have not preserved full momentum resolution \citep{suenaga2000element,suenaga2009visualizing,ramasse2013probing,lin2015exploring}. Nevertheless, we show that the stringent count requirements can be substantially relaxed through statistical inference, judicious selection of the ionization edge and a favourable signal-to-background ratio (SBR).
% ,  suggesting that momentum-resolved, single-atom depth localization may be achievable at experimentally accessible electron doses.

\section{Core-Loss 4D-STEM Interpretation in the Born Approximation}

Core-loss scattering can be implemented within a multislice framework  \citep{dwyer2005multislice} wherein it is assumed that the entire core-loss interaction is confined to a single slice. Consequently, prior to a core-loss interaction, the wavefield is propagated elastically. At the plane of ionization, an inelastic wavefield, $\psi_f$, is derived from the incoming elastic wavefield, $\psi_i$, via a  multiplicative transition potential, $H_{fi}$, 
\begin{align}
    \psi_f = -i \sigma_{fi} H_{fi} \psi_i.
    \label{eq:Hpsi}
\end{align}
where $\sigma_{fi}$ is the usual interaction parameter with wave vector corresponding to the beam electron after the ionization event.
Since a spectrometer filters based only on energy loss, the spectrometer is blind to the specific mechanism that transpired. Consequently, every ionization pathway yielding the requisite energy deficit must be accounted for. 
Each pathway is represented by a distinct transition potential, which encodes a specific core-loss ionization mechanism. 
In the present treatment, these degenerate ionization pathways are parameterized by the spin and magnetic quantum number of the initially bound core electron, together with the continuum state of the ejected electron. Analogously to bound-state orbitals, the continuum states are characterized by orbital and magnetic quantum numbers, as well as the energy above the ionization threshold. 

Following ionization, each inelastic wave, corresponding to a particular combination of the aforementioned degrees of freedom, is propagated elastically from the ionization site to the exit surface using the multislice algorithm. The final 4D-STEM intensity distribution is given by the incoherent sum over all ionization pathways and all possible depths at which ionization may occur
\begin{equation}
I(\mathbf{R}, \mathbf{q}) = \sum_{fi} \Bigg| \mathcal{F} \Bigl\{ \mathcal{M}_{z' \rightarrow \mathrm{exit}} \,(-i\sigma_{fi}) H_{fi} \mathcal{M}_{\mathrm{entrance} \rightarrow z'} \, \psi_0(\mathbf{r} - \mathbf{R}) \Bigr\} \Bigg|^2,
\label{eq:I_final}
\end{equation}
where $\mathcal{M}$ denotes the multislice propagation operator, acting on all quantities to its right. Here $z'$ is the ionization depth measured from the specimen entrance surface and $\psi_0$ represents the incident probe wavefunction centred at scan position $\mathbf{R}$. The sum over $fi$ encompasses all ionization pathways and atomic sites capable of contributing to the selected energy-loss. 
% Furthermore, the incoherent summation must extend over all atomic sites of the relevant elemental species capable of contributing to the selected energy-loss window.

\begin{figure*}
    \centering
    \includegraphics[width=1\linewidth]{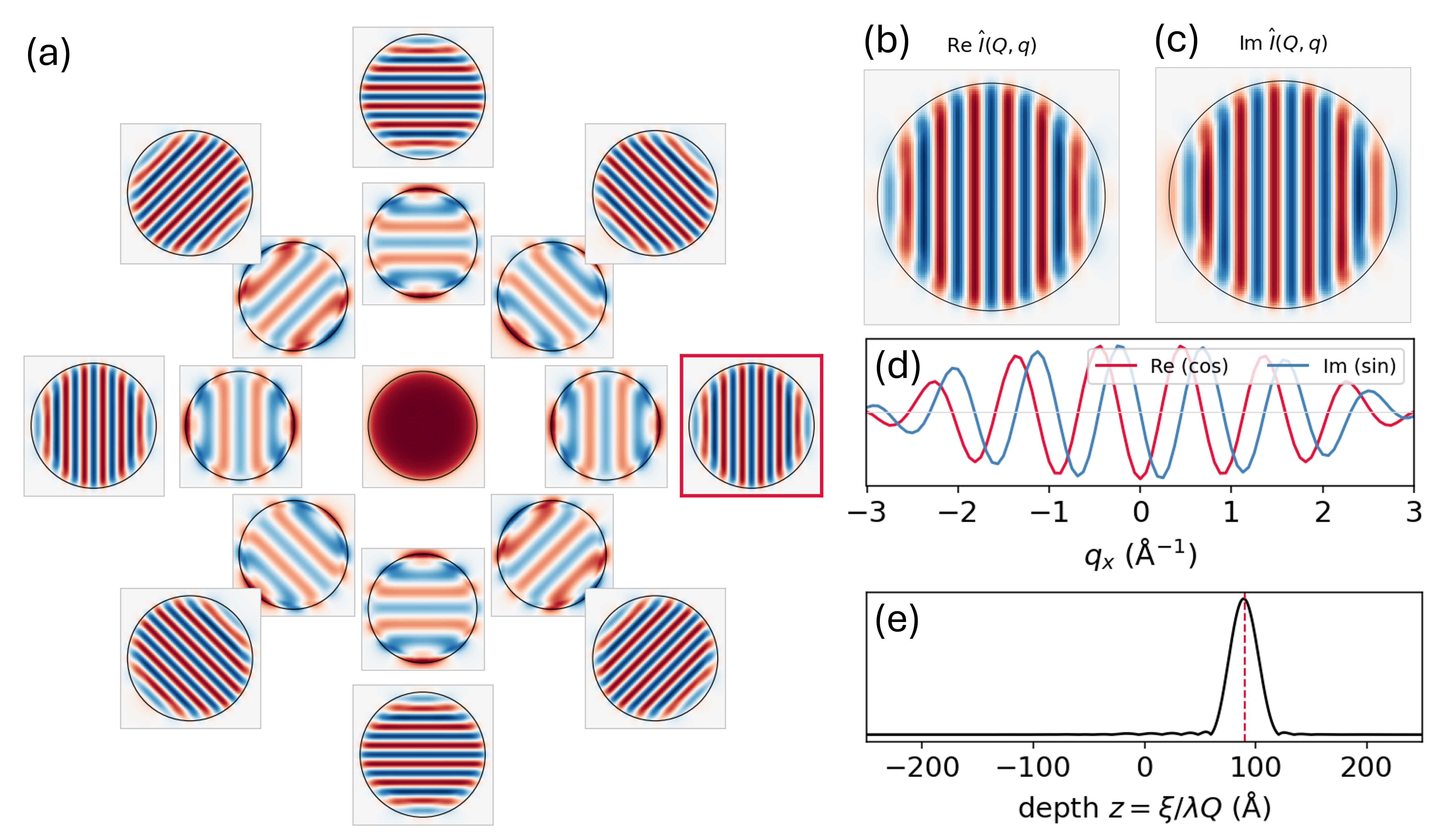}
    \caption{(a) The real component of a Fourier transformed core-loss 4D-STEM dataset, $\text{Re}\{\hat I(\vQ,\vq)\}$, wrt the probe coordinate, $\vQ$, arranged according to magnitude and azimuthal angle of $\vQ$. Patterns exhibit fringes with frequency according to the atom’s distance along the optic axis from the focal plane.  The (b) real and (c) imaginary components of the framed pattern in (a) show fringes in both components. (d) shows the real and imaginary components projected along the length of the fringes indicating both components have a common frequency but a relative phase shift. (e) shows the Fourier transform of the complex projected fringes (with rescaled axis) with peak located at the iron atom’s distance of 90 $\AA$ from the focal plane. The fringe profile is formed by averaging detector pixels within bins along \(\hat{\mathbf Q}\), then applying a Hann window spanning the bright-field diameter to suppress spectral leakage. Simulation assumes 60 mrad probe, 300 keV beam energy and Fe-L$_{2,3}$ edge is utilized for core-loss signal.}
    \label{fig:mosaic}
\end{figure*}

To elucidate the information contained in core-loss 4D-STEM data, we specialize to the Born approximation, which assumes all elastic propagation is through vacuum. We will relax this assumption in \cref{sec:multiple_scattering} to explore the neglected effects of multiple scattering. 
% Within the Born approximation, the multislice operator reduces to a Fresnel propagation. 
For an atom suspended in vacuum, Fourier transforming \cref{eq:I_final} over the probe coordinate, $\vR \rightarrow \vQ$, obtains (see derivation in \cref{sec:appendix_IQq_derivation})
\begin{align}
    \hat{I}(\mathbf{Q},\mathbf{q}) 
    &= \int I(\mathbf{R},\mathbf{q})\,
       e^{-2\pi i \mathbf{Q}\cdot\mathbf{R}}\,d^2R \nonumber\\
    &=   e^{-2\pi i \mathbf{Q}\cdot\mathbf{x}}\;
                  e^{\,2 i\pi\lambda z \mathbf{q}\cdot\mathbf{Q}} \nonumber\\
    &\quad\times \int d^2u\;
      \sum_{fi} \sigma_{fi}^2 \hat{H}_{fi}\!\left(\mathbf{u}+\tfrac{\mathbf{Q}}{2}\right)
      \hat{H}_{fi}^*\!\left(\mathbf{u}-\tfrac{\mathbf{Q}}{2}\right) \nonumber\\
    &\qquad\times
      \hat{\psi}_0\!\left(\mathbf{q}-\mathbf{u}-\tfrac{\mathbf{Q}}{2}\right)
      \hat{\psi}_0^*\!\left(\mathbf{q}-\mathbf{u}+\tfrac{\mathbf{Q}}{2}\right) \nonumber
     \\ &\qquad\times e^{-2\pi i\lambda z\,\mathbf{u}\cdot\mathbf{Q}}.
    \label{eq:I_Qq}
\end{align}
where $\hat \psi_0$ denotes the reciprocal space probe wavefunction at the focal plane and $z=z'+\Delta f$ is the distance between the ionization event and focal plane, with underfocus defined as negative.
The transmission function of the ionized atom has been omitted for simplicity, as its inclusion would complicate subsequent analysis while introducing only minor corrections.
However, at any time it may be reinserted by substituting $H \rightarrow tH$ or $\hat H \rightarrow \hat t \conv \hat H$, where $\conv$ denotes a convolution.
\Cref{fig:mosaic}(a) shows an example of \cref{eq:I_Qq} for a single Fe atom suspended in vacuum, simulated with a 300 keV probe with 60 mrad probe-forming aperture semi-opening angle. The Fe-L$_{2,3}$ edge has been selected for its delocalized interaction, which we will find is favourable for dopant depth retrieval. The figure presents a curated subset of the resulting Fourier transformed 4D-STEM dataset, with  diffraction patterns arranged according to the magnitude and azimuthal angle of the probe-position reciprocal-space coordinate, $\vQ$.  A striking pattern of fringes emerges where the oscillation frequency is dictated by $Q$, the radial distance from the centre of the mosaic, while the fringes orientation is governed by $\hat\vQ$, the azimuthal angle around the mosaic. The central tile, corresponding to $Q=0$, is a core-loss position-averaged convergent-beam electron diffraction (PACBED) pattern which, in the absence of channelling, admits no structure  \citep{deimetry2025dopant}. 
Since \cref{eq:I_Qq} is generally complex, only the real component of the diffraction pattern is presented. 
The real and imaginary pair for a single pattern, highlighted with a red border in \cref{fig:mosaic}(a), is shown in  \cref{fig:mosaic}(b) and (c), respectively, revealing the same fringe structure is common to both. If the fringes are collapsed along the longitudinal axis, shown in \cref{fig:mosaic}(d), it becomes clear the oscillation frequency is the same but the components are phase-shifted relative to one another.

Returning to \cref{eq:I_Qq}, these fringes originate from the phase prefactor $e^{\,2 i\pi\lambda z \mathbf{q}\cdot\mathbf{Q}}$. The oscillation frequency is given by $\lambda z \vq \cdot \vQ$ and varies with the relative angle between the detector and probe position reciprocal space coordinates. For fixed $\vQ$, when $\vq$ runs parallel to $\vQ$, the oscillation frequency reduces to $\lambda z q Q$. Conversely, when $\vq$ runs perpendicular to $\vQ$, the oscillation frequency reduces to zero. 
Because the phase depends on $\mathbf{q}$ only through its projection onto $\mathbf{Q}$, the prefactor is invariant under displacements of $\mathbf{q}$ perpendicular to $\mathbf{Q}$ so the fringes run perpendicular to $\vQ$, tracing contours of constant $\mathbf{q}\cdot\mathbf{Q}$. This explains the radial and azimuthal dependence on $\vQ$ seen in \cref{fig:mosaic}(a). Additionally, we see that the oscillation frequency is dictated by the beam electron wavelength and the iron atom's depth $z$. This implies that when studying the fringe oscillation frequency, the axis should be rescaled by $\lambda Q$ to directly correspond with the atom's depth. The Fourier spectrum for the phasor with components in \cref{fig:mosaic}(d) is shown in \cref{fig:mosaic}(e) with the axes rescaled to atomic depth, indicating the atom is 90${\rm\AA}$ from the focal plane. 
The phase shift of the imaginary component relative to the real component is contingent upon the side of the focal plane on which the atom resides, a consequence of the sign of $z$ within the prefactor. 
% The peak in \cref{fig:mosaic}(e) is consistent with the convention adopted here, wherein the atom lies further along the optic axis than the probe cross-over point ($z > 0$). 
Were the atom instead situated on the opposite side of the focal plane, such that $z < 0$, the imaginary component would acquire an additional $180^\circ$ phase shift, manifesting in the Fourier spectrum as a peak reflected to the mirror-symmetric depth.

The additional integral term in \cref{eq:I_Qq}, which modulates the complex exponential responsible for the fringe contrast, introduces several complications to the preceding interpretation. First, the fringes are predominantly confined to the bright-field disk, motivating an average rather than a sum when collapsing the fringes to account for their varying lengths across the bright-field region. Second, the modulation introduced by the probe and transition potentials can give rise to spectral artefacts, including spectral leakage. To mitigate these effects, a Hann window is applied over the bright field region following fringe collapse. These alterations have been applied to \cref{fig:mosaic}(e).

% A coordinate system which more naturally accommodates the fringes is one in which all vector quantities are decomposed according to a component parallel and perpendicular to $\vQ$. 

An analytic expression for the spectrum of \cref{fig:mosaic}(e) can be obtained by recognizing that \cref{eq:I_Qq} can be rewritten as a convolution 
\begin{equation}
\hat{I}(\vQ,\vq) = e^{-2\pi i \vQ\cdot {\bf x}} \Bigl[\hat{\mathcal{H}}(\vQ, \vv) \conv_{\vv} \bigl(\hat{\mathcal{P}}(\vQ, \vv)\,e^{2\pi i \lambda z \vQ \cdot \vv}\bigr)\Bigr](\vq),
\label{eq:gauge}
\end{equation}
with definitions
\begin{samepage}
\begin{equation}
\hat {\mathcal{H}}(\vQ,\vv) = \sum_{fi} \sigma_{fi}^2 \hat{H}_{fi}\!\bigl(\vv+\tfrac{\vQ}{2}\bigr)\,
                      \hat{H}_{fi}^{*}\!\bigl(\vv-\tfrac{\vQ}{2}\bigr), \label{eq:Hn0_AF_qspace}
\end{equation}
\begin{equation}
\hat {\mathcal{P}}(\vQ,\vv) = \hat{\psi}_0\!\bigl(\vv-\tfrac{\vQ}{2}\bigr)\,
                      \hat{\psi}_0^{*}\!\bigl(\vv+\tfrac{\vQ}{2}\bigr). \label{eq:probe_AF_qspace}
\end{equation}
\end{samepage}
The prefactor $e^{\,2 i\pi\lambda z \mathbf{q}\cdot\mathbf{Q}}$ has cancelled and now appears as a phase factor associated with $\hat {\mathcal{P}}$. 
Forward Fourier transforming \cref{eq:gauge} over $\vq \to \vxi$ \footnote{The convention of whether to use a forward or inverse Fourier transform to obtain $\vq \to \vxi$ only affects the sign convention used to specify dopant's depth relative to the focal plane.}, the convolution reduces to a simple product,
\begin{align}
    \hat{I}(\vQ,\vxi) &=  e^{-2\pi i \vQ \cdot \vx}\,\hat{\mathcal{H}}(\vQ,\vxi)\,\hat{\mathcal{P}}\bigl(\vQ,\vxi - \lambda z Q \hat\vQ\bigr),
\label{eq:I_FT}
\end{align}
where $\vxi$ carries dimensions of length, and
\begin{samepage}
\begin{equation}
\hat{\mathcal{H}}(\vQ,\vxi) = \mathcal{F}_{\vv\rightarrow\vxi}\{\hat{\mathcal{H}}(\vQ,\vv)\} = \int d^2\chi\; \sum_{fi} \sigma_{fi}^2 {H}_{fi}\!\Bigl(\vchi-\tfrac{\vxi}{2}\Bigr) {H}_{fi}^{*}\!\Bigl(\vchi+\tfrac{\vxi}{2}\Bigr)\, e^{-2\pi i \vQ\cdot\vchi}, \label{eq:hn0_amb}
\end{equation}
\begin{equation}
\hat{\mathcal{P}}(\vQ,\vxi) = \mathcal{F}_{\vv\rightarrow\vxi}\{\hat{\mathcal{P}}(\vQ,\vv)\} = \int d^2\chi\; {\psi}_0\!\Bigl(\vchi-\tfrac{\vxi}{2}\Bigr) {\psi}_0^{*}\!\Bigl(\vchi+\tfrac{\vxi}{2}\Bigr)\, e^{+2\pi i \vQ\cdot\vchi} \label{eq:probe_amb}
\end{equation}
\end{samepage}
which are, respectively, the ambiguity functions of the transition potential and the probe, both related to their corresponding Wigner distributions by a Fourier transform. \Cref{eq:I_FT} will be dubbed a depth spectrum despite the coordinate no longer being a reciprocal of length. This nomenclature has been chosen since \cref{fig:mosaic}(e) resembles a Fourier spectrum and was obtained via a forward transform.

\section{Structure of the Probe and Transition Potentials Ambiguity Functions}

\begin{figure*}
    \centering
    \includegraphics[width=1\linewidth]{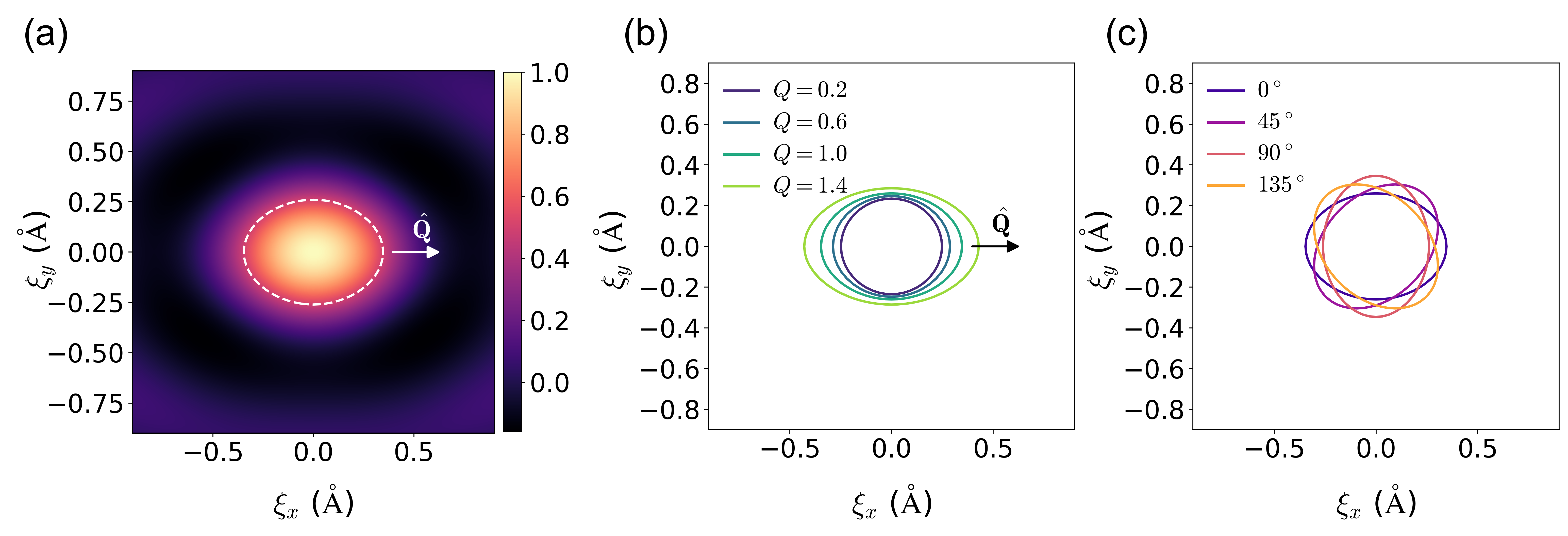}
    \caption{(a) Aberration-free 30 mrad normalized probe ambiguity function as defined in \cref{eq:probe_amb} for $\vQ=(1, 0) \; {\rm \AA}^{-1}$. (b) shows probe ambiguity function contour at half-maximum which elongates as magnitude of $\vQ$ grows. (c) shows probe ambiguity function contour at half-maximum with $Q=1\;{\rm \AA}^{-1}$ fixed while $\hat\vQ$ changes direction. The ambiguity function now elongates in the direction of $\hat\vQ$.}
    \label{fig:probe_af_2d}
\end{figure*}

\Cref{eq:I_FT}  reveals that the entire 4D core-loss signal, two dimensions carried by $\vQ$ which is the spatial frequency conjugate to the probe position, and two carried by $\vxi$ which is the length-dimensioned variable conjugate to the detector coordinate $\vq$, factorizes into the product of two ambiguity functions, one intrinsic to the transition potential and one intrinsic to the probe. This factorization simplifies the interpretation of the depth spectrum which can be fully elucidated by a thorough understanding of the individual ambiguity functions. 
 The transition potentials ambiguity function, $\hat{\mathcal{H}}(\mathbf{Q},\boldsymbol{\xi})$, remains anchored at the origin of $\boldsymbol{\xi}$-space irrespective of $\mathbf{Q}$. Its footprint reflects only the internal structure of the ionization event i.e. the atomic species, edge, bound and continuum state wavefunctions etc., and carries no memory of where along the optic axis the core-loss event took place. Conversely, the probe ambiguity function, $\hat{\mathcal{P}}(\mathbf{Q},\boldsymbol{\xi})$,  is evaluated at the shifted argument $\boldsymbol{\xi}-\lambda z Q\hat{\mathbf{Q}}$  along the direction $\hat{\mathbf{Q}}$. This displacement crucially contains the depth $z$ at which the atom resides relative to the probe's focal plane. 
% Therefore, characterization of the probe ambiguity function's shape is sufficient to anticipate the form of the depth spectrum for any choice of $\mathbf{Q}$. 
\Cref{fig:probe_af_2d}(a) presents the probe ambiguity function evaluated at $\mathbf{Q}=(1,0)$\AA$^{-1}$, displayed as a function of $\boldsymbol{\xi}=(\xi_x,\xi_y)$. Although the aberration-free probe used is rotationally symmetric, the resulting ambiguity function is visibly elongated along $\hat{\mathbf{Q}}$, which is made explicit by the dashed contour drawn at half the peak amplitude. This elongation intensifies as $ Q$ grows as shown in \cref{fig:probe_af_2d}(b) which shows only the half-maximum contour for several increasing values of $ Q$ while holding $\hat{\mathbf{Q}}$ fixed. As $ Q$ grows, the eccentricity of the ambiguity function grows monotonically in the direction of $\hat{\vQ}$.
\Cref{fig:probe_af_2d}(c) instead holds $ Q$ fixed while sweeping the azimuthal direction $\hat{\mathbf{Q}}$. Now the contour simply rotates, tracking $\hat{\mathbf{Q}}$ while preserving its shape and eccentricity. 
Neither of these behaviours arises by coincidence. For a probe with only radially symmetric aberrations,  the ambiguity function can depend only on $ \xi$, $ Q$, and the relative angle between $\boldsymbol{\xi}$ and $\mathbf{Q}$. 
Hence, in \cref{fig:probe_af_2d}(c), a coordinate transformation of $\boldsymbol{\xi}$ into components parallel and perpendicular to $\hat{\mathbf{Q}}$  is indistinguishable from the $0^\circ$ contour.

% todo: Hn0 ambiguity function\cite{\cite{\cite{}}}

\begin{figure}
    \centering
    \includegraphics[width=1\linewidth]{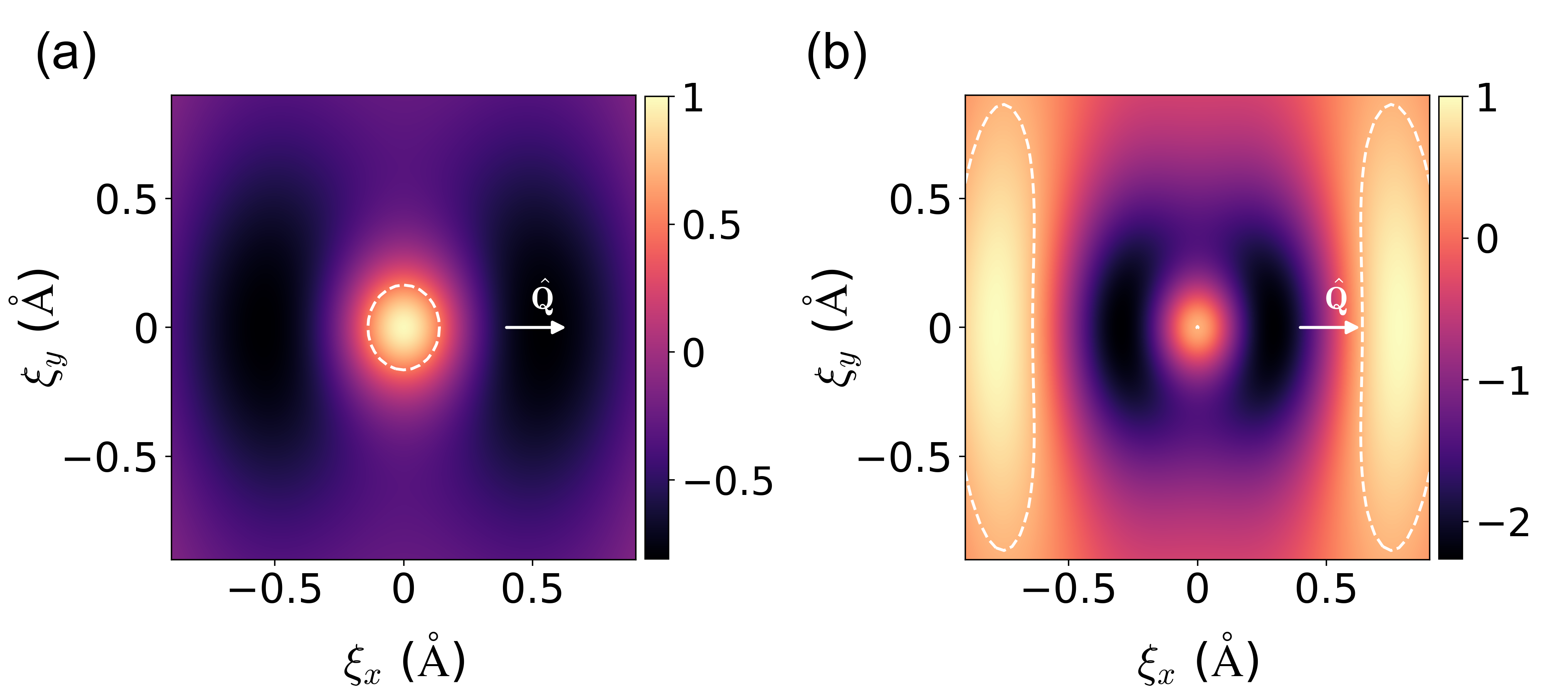}
    \caption{Real component of Fe-L$_{2,3}$ normalized transition potentials ambiguity function as defined in \cref{eq:hn0_amb} for (a) $\vQ=(1, 0) \; {\rm \AA^{-1}}$ and (b) $\vQ=(2, 0) \; {\rm \AA^{-1}}$. Unlike the probe ambiguity function, the transition potentials ambiguity function compresses with increasing $Q$.
}
    \label{fig:Hn0_examples}
\end{figure}

\begin{figure}
    \centering
    \includegraphics[width=1\linewidth]{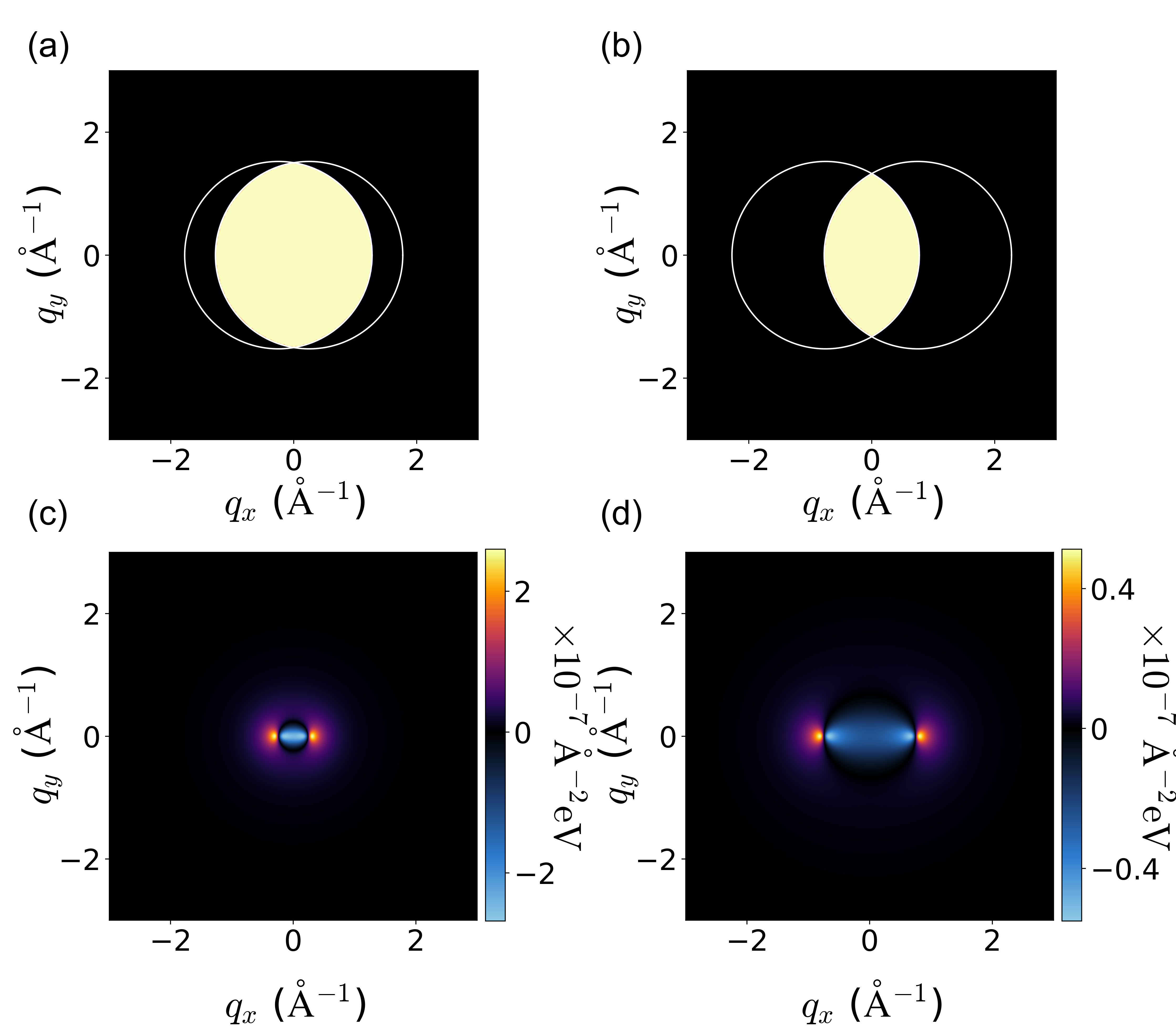}
    \caption{Aberration-free 30 mrad, 300 keV inverse Fourier transform of probe ambiguity function as specified in \cref{eq:probe_AF_qspace} for (a) $Q=0.5\; {\rm \AA^{-1}}$ and (b) $Q=1.5 \; {\rm \AA^{-1}}$. Increasing $Q$ relatively elongates the overlapping region along $q_y$. 
    In contrast, the real component of the inverse Fourier transform of the Fe-L$_{2,3}$ transition potentials ambiguity function, as defined in \cref{eq:Hn0_AF_qspace}, for (c) $Q=0.5\;{\rm \AA^{-1}}$ and (d) $Q=1.5\;{\rm \AA^{-1}}$, shows the overlapping region elongating along $q_x$ as $Q$ increases. Transitions calculated up to second order and 10 eV above ionization threshold. 
}
    \label{fig:gfunc_ex}
\end{figure}

% The transition potentials ambiguity functions abides by similar rules. 
Surprisingly, like the probe ambiguity function, the transition potentials ambiguity function also only depends on $ Q$, $ \xi$ and the angle between  $\vQ$ and $\vxi$, which is proven in \cref{sec:Hn0_iso_proof} for any edge using the  parameterization of \cite{dwyer2005multislice}. 
Shown in \cref{fig:Hn0_examples} is the Fe-L$_{2,3}$ edge ambiguity function for (a) $Q=1\;{\rm \AA}^{-1}$ and (b) $Q=2\;{\rm \AA}^{-1}$. While previously we saw that the probe ambiguity function elongated as $Q$ increased in magnitude, now we see the opposite where the ambiguity function appears to contract. 
The reason is made clear by studying \cref{eq:Hn0_AF_qspace,eq:probe_AF_qspace} directly. 
\Cref{fig:gfunc_ex}(a) and (b) show \cref{eq:probe_AF_qspace} for (a) $Q=0.5\;{\rm \AA}^{-1}$ and (b) $Q=1.5\mathrm{\AA^{-1}}$ corresponding to the overlap between two copies of the aperture function separated by $Q$.
Expectedly, the overlap region narrows more quickly in the direction of $\hat\vQ$ than in the direction of $\hat\vQ_\perp$ causing the overlap to effectively elongate along $\vQ_\perp$. As such, the behaviour in the Fourier domain will be swapped as we have concluded from \cref{fig:probe_af_2d}(b).
Similarly, shown in \cref{fig:gfunc_ex}(c) and (d) is \cref{eq:Hn0_AF_qspace} for (c) $Q=0.5\;{\rm \AA}^{-1}$ and (d) $Q=1.5\;{\rm \AA^{-1}}$. Now, moving the transition potentials further apart widens $\hat{\mathcal{H}}(\vQ,\vq)$ in the direction of $\hat\vQ$, resulting in behaviour contrary to that observed for the probe ambiguity function in \cref{fig:probe_af_2d}.
% This implies that as $Q$ increases, the depth spectrum peak will shift further from the origin (due to the increasing fringe frequency), the peak will elongate reducing depth resolution and the transition potential will become increasingly localized suppressed the probe ambiguity function i.e. suppressing the depth spectrum peak.  

These findings imply the 4D-space in which the dataset is represented is redundant. The depth information we ultimately seek is captured by how far the probe ambiguity function is displaced from the origin. Hence, depth information is not distributed isotropically, but most in the $\hat\vQ$ direction. 
Both ambiguity functions follow the angle of $\hat\vQ$ implying the content of interest within the full $\boldsymbol{\xi}$-plane is effectively one-dimensional. Thus, a single line scan through the origin, taken along $\hat{\mathbf{Q}}$, is sufficient to locate the peak and thus recover $z$. The remaining dimension of $\boldsymbol{\xi}$-space, not parallel with $\hat{\mathbf{Q}}$, is negligible for this purpose.
Translated back into the conjugate detector coordinate $\mathbf{q}$, the proposed line scan corresponds to collapsing the diffraction pattern in the direction perpendicular to $\mathbf{Q}$ which is precisely the operation of collapsing the fringe pattern down into a 1D spectrum as done from \cref{fig:mosaic}(b) and (c) to (d). The spectrum shown in \cref{fig:mosaic}(e) is therefore the envelope of the probe ambiguity function, in \cref{eq:probe_amb}, along $\hat\vQ$ and modulated by the transition potentials ambiguity function (and convolved with windowing and normalization functions).
% This superfluity was first hinted when studying \cref{fig:mosaic}(a) where we concluded that the oscillation frequency of the fringes, which are of primary importance, are most effectively captured by sampling along  $\hat\vQ$.
% Collapsing the fringes therefore distills the full 4D-STEM dataset to maximize the depth information of interest. 
Additionally, collapsing the fringe pattern enhances the signal-to-noise ratio, while restricting the analysis to a profile along $\hat\vQ$ removes sensitivity to structure in the $\vxi$-plane that is affected by departures from the idealizations of the Born approximation without providing additional depth information.

Motivated by the preceding argument, we restrict our attention to the ambiguity functions evaluated along $\xi_\parallel = \boldsymbol{\xi}\cdot\hat{\mathbf{Q}}$, that is, the line scan passing through the origin and through the probe ambiguity function peak identified above. 
The reduced dimensionality allows us to better visualize the functional dependence of the ambiguity functions with $\vQ$ which will be vital for studying how depth information is encoded in different regions of the 4D-STEM dataset. 
 
\begin{figure}
    \centering
    \includegraphics[width=1\linewidth]{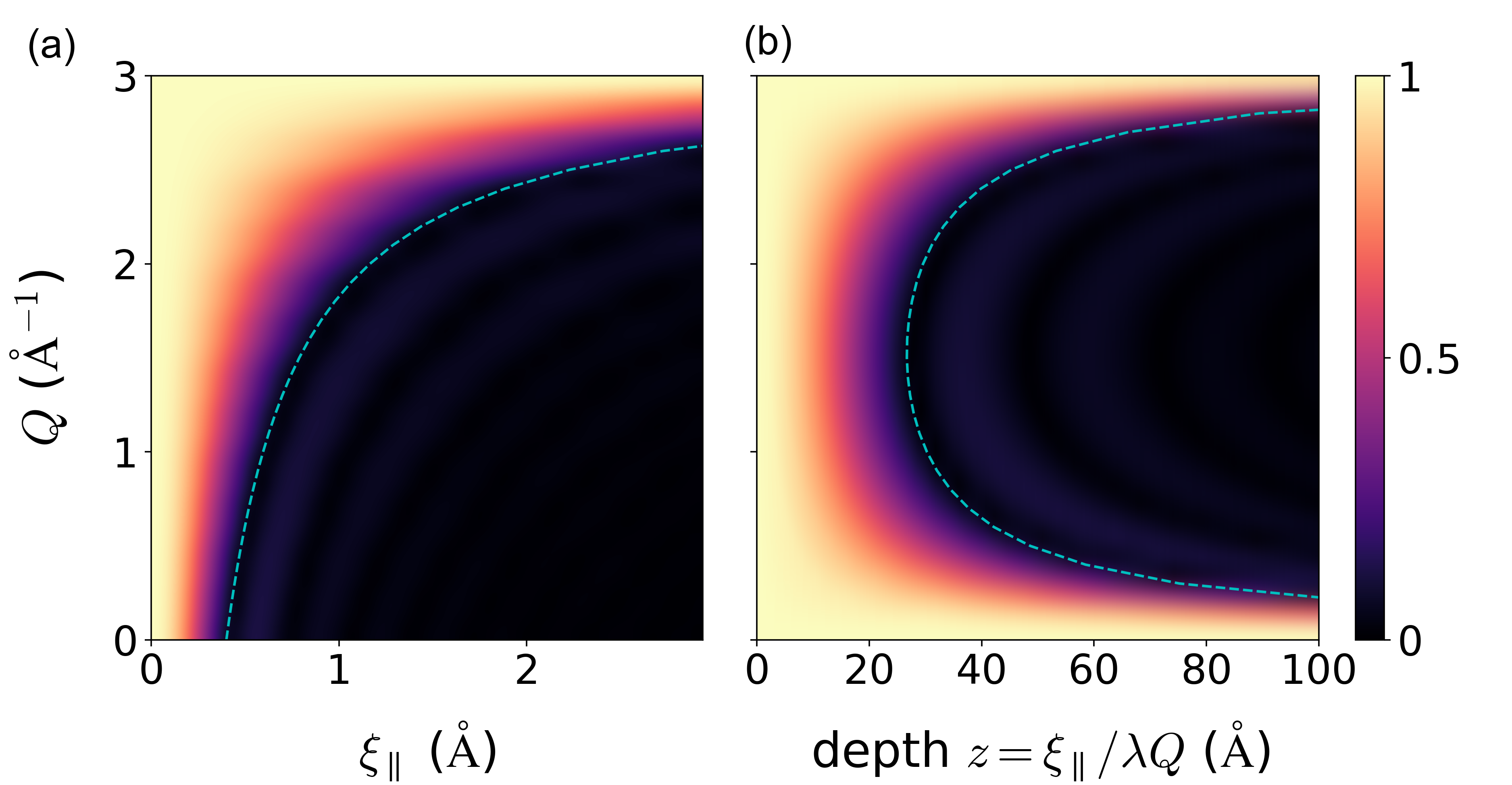}
    \caption{Probe ambiguity function as a function of $\vQ=Q\hat{\vQ}$ and (a) $\xi_\parallel=\vxi\cdot\hat{\vQ}$ and (b) atom depth. Depth-resolving power is optimal at $Q=q_{\rm max}$, which balances probe overlap against fringe frequency. Each row of constant $Q$ is normalized by its maximum to remove $Q$-dependent intensity variations. The probe is aberration-free, with a 30 mrad convergence semi-angle and 300 keV accelerating voltage.}
    \label{fig:probe_ambig_fcns}
\end{figure}

 \Cref{fig:probe_ambig_fcns}(a) shows the probe ambiguity function, $\hat{\mathcal{P}}(Q,\xi_\parallel)$. Consistent with the elongation established in \cref{fig:probe_af_2d}, its width along $\xi_\parallel$ grows monotonically with increasing $Q$. This can be quantified via the Rayleigh criterion in terms of the first zero-crossing. While the first zero-crossing cannot be obtained analytically for $Q\neq0$, we find empirically that 
 \begin{equation} \xi_0(Q)\;\simeq\;\frac{1.22}{2q_{\max}-Q} \label{eq:zc} \end{equation} 
where $q_{\max}=\alpha/\lambda$ is the aperture radius in reciprocal space, $\alpha$ the semi-convergence angle and $\lambda$ the relativistically corrected wavelength. \Cref{eq:zc} is overlaid on \cref{fig:probe_ambig_fcns}(a) as a dotted line and is indeed in good agreement. Using $\xi=Q\lambda z$, the Rayleigh criterion in terms of the dopant depth is given by
% Now we wish to determine the resolving power in coordinates of $\xi=Q\lambda z$, the Rayleigh criterion for coordinate $\xi$ is translated to $z$ by
 \begin{align} \Delta z_{\mathrm{Rayleigh}} = \frac{\xi_0}{\lambda Q} \;\simeq\;\frac{1.22}{\lambda\,Q\,(2q_{\max}-Q)} = \frac{1.22}{Q(2\alpha - \lambda Q)}. \label{eq:zrayleigh} \end{align} 
The factor \(2q_{\max}-Q\) is the overlap, parallel to \(\hat\vQ\), between two circular probe-forming apertures whose centres are separated by \(Q\). 
The probe ambiguity function is again shown in \cref{fig:probe_ambig_fcns}(b) as a function of $z$, rather than $\xi$, to directly elucidate how the width of probe ambiguity function scales with dopant depth. 
% Since depth resolution is governed by the probe ambiguity function, \cref{fig:probe_ambig_fcns}(b) shows a 30 mrad probe ambiguity function as a function of $z = \xi_\parallel / (\lambda Q)$, rather than $\xi_\parallel$. 
The depth resolution is visibly lost at both edges of \cref{fig:probe_ambig_fcns}(b), near $Q \sim 0$ and $Q \sim 2q_{\rm max} \approx 3{\rm \AA^{-1}}$. This follows from the depth resolution being governed by a competition between the increasing depth sensitivity $\lambda Q$ and the decreasing pupil overlap width $2q_{\max} - Q$. As $Q \to 0$ the depth-dependent fringe frequency vanishes, and as $Q \to 2q_{\max}$ the pupil overlap vanishes, so resolving power deteriorates at both extremes. \Cref{fig:probe_ambig_fcns}(b) implies the ambiguity function is sharpest near $Q_{\rm opt} \approx q_{\max} \approx 1.5{\rm \AA^{-1}}$, consistent with the minimum of \cref{eq:zrayleigh} which corresponds to 
% At $Q_{\rm opt}$
\begin{align}
    \Delta z_{\min}
    \approx \frac{1.22\lambda}{\alpha^2}.
    \label{eq:zmin}
\end{align}
This implies that axial resolution improves with increasing aperture size and accelerating voltage.
This shares the $\lambda/\alpha^2$ scaling of the conventional depth of field of an Airy probe  \citep{born1999principles} which has been observed in a range of contexts  \citep{borisevich2006depth,deimetry2026threedimensionaldopantdistributionsingle,lupini2011three}, while the numerical prefactors depend on the width-specific criterion chosen. 

\begin{figure}
    \centering
    \includegraphics[width=1\linewidth]{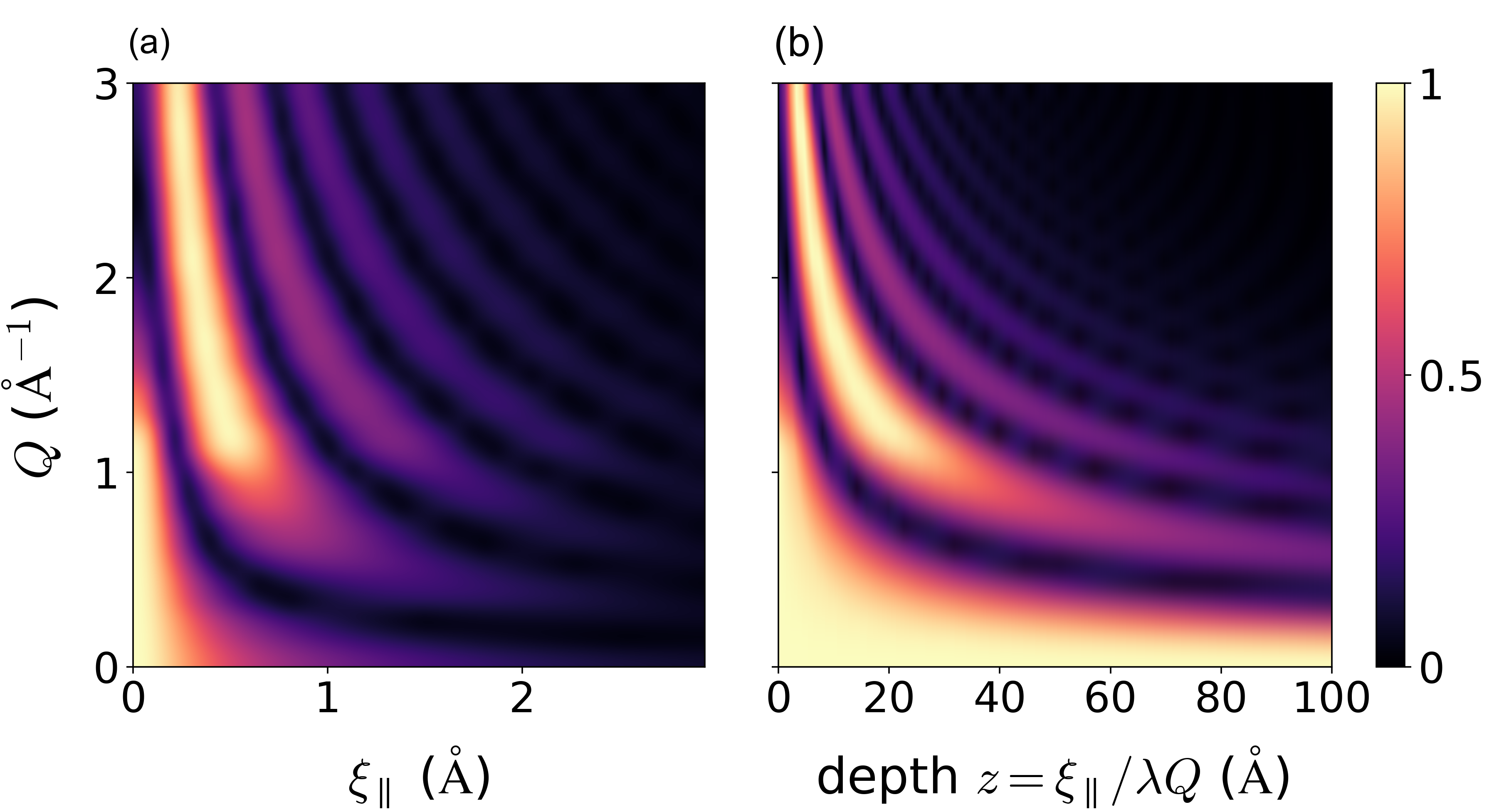}
    \caption{Magnitude of Fe-L$_{2,3}$ transition potentials ambiguity function as a function of $\vQ=Q\hat{\vQ}$ and (a) $\xi_\parallel=\vxi\cdot\hat{\vQ}$ and (b) atom depth. Each row of constant $Q$ is normalized by its maximum to remove $Q$-dependent intensity variations. Transition potentials calculated at 10 eV above ionization threshold.}
    \label{fig:hn0_ambig_fcns}
\end{figure}

Similarly, the transition potentials ambiguity function along $\hat\vQ$ is shown as a function of $Q$ and $\xi_\parallel$ in \cref{fig:hn0_ambig_fcns}(a). At $Q=0$, the ambiguity function extends throughout most of the $\xi_\parallel$ range shown.
At small $Q$, the function varies slowly with $\xi_\parallel$, whereas increasing $Q$ produces a series of lobes with nodes in-between.
If the depth spectrum peak happens to coincide with a node, then a significant suppression will result. However, since the probe ambiguity function displacement and transition potentials ambiguity function nodes vary with $Q$, the peak will not consistently coincide with a single node. As such, some patterns will retain more depth information than others.
For comparison with \cref{fig:probe_ambig_fcns}(b), the transition potentials ambiguity function is also shown in \cref{fig:hn0_ambig_fcns}(b) as a function of depth. Unlike the probe ambiguity function, no particularly notable changes in structure arise. 

Having elucidated the characteristics of the depth spectrum, we briefly turn to the properties that determine the suitability of a core-loss transition for this technique.
Since the depth spectrum in \cref{eq:I_FT} is given by the product of the transition potential and probe ambiguity functions, ideally, the transition potentials ambiguity function should vary slowly and extend over a broad range in $\vxi$-space. This is desirable for several reasons. 
First, a slowly varying transition potentials ambiguity function introduces minimal modulation across the support of the probe ambiguity function i.e. the transition potentials ambiguity function can be well approximated as $\xi$-independent over the width of the probe ambiguity function.
Second, a broader transition potentials ambiguity function accommodates larger probe ambiguity function displacements from the origin without significantly suppressing the signal, thereby extending the accessible depth range.
Third, as seen in \cref{fig:Hn0_examples}, increasing $Q$ towards $Q_{\rm opt}$ contracts the transition potentials ambiguity function, progressively suppressing depth information even as resolving power improves. 
% Therefore, a slower contraction is desirable to best access the range of optimal depth resolution.
A spatially delocalized transitions is broad in real space, which produces a narrow Fourier-space distribution and hence a narrow overlap in \cref{eq:Hn0_AF_qspace}, which, upon Fourier transformation again, yields a broad ambiguity function in $\vxi$-space.

\section{Channelling}\label{sec:multiple_scattering}

To this point, dynamical scattering has been avoided by studying an isolated atom in vacuum, where electron propagation is analytically tractable. 
While the ionization event itself redistributes the electron momentum, additional redistribution arises from elastic scattering by the host crystal before and after ionization. These additional interactions are absent from the Born approximation and may compromise the depth spectrum interpretation developed above.

\begin{figure}
    \centering
    \includegraphics[width=1\linewidth]{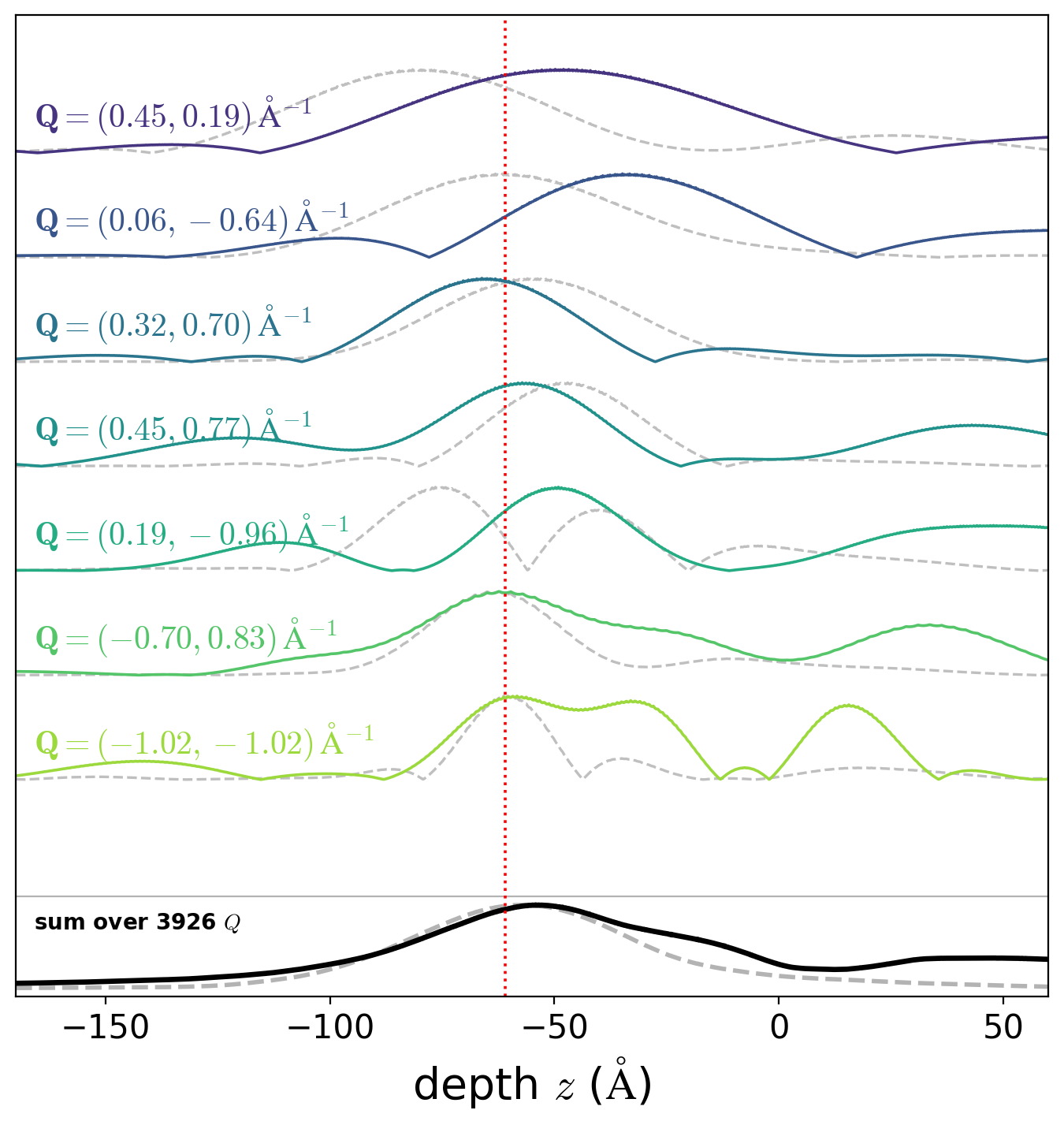}
    \caption{
Single-$\mathbf{Q}$ depth spectra of the Fe-L$_{2,3}$ signal for an iron atom in vacuum (grey dashed) and an iron dopant $39.1\;{\rm \AA}$ deep in $20\;\mathrm{nm}$ of [001] SrTiO$_3$ (coloured), at $-60.9\;{\rm \AA}$ from focus (red dotted line). Individual $\mathbf{Q}$ spectra, which show spurious features, are suppressed when averaged over $0.4\leq Q\leq 2.7\;{\rm \AA}^{-1}$ (bottom), only in the case of vacuum-suspended iron atom where dynamical scattering is not present. The probe convergence semi-angle is \(30\mathrm{ mrad}\), and the accelerating voltage is \(300\;\mathrm{keV}\). The fringe profile is formed by averaging detector pixels within bins along \(\hat{\mathbf Q}\), then applying a Hann window spanning the bright-field diameter to suppress spectral leakage.}
    \label{fig:Q-indiv}
\end{figure}

Consider \cref{fig:Q-indiv}, which presents a selection of depth spectra at various $\vQ$, shown in dashed grey,  for an iron atom suspended in vacuum $-61\rm \AA$ from the focal plane.
Despite the absence of dynamical scattering, confounding spectral features are still present due to the modulation of the displaced probe ambiguity function by the transition potentials ambiguity function.
Fortunately, these features are inconsistent across $\vQ$, suggesting that their influence can be suppressed via averaging. This reinforces features consistent with the dopant depth while suppressing spurious sporadic contributions. 
Specifically, we adopt a scheme wherein the magnitude of each depth spectrum is interpolated onto a common depth axis and averaged across $\vQ$.
The depth spectra averaged over the range $0.4 \leq  Q \leq 2.7\;{\rm Å}^{-1}$ for the vacuum-suspended iron atom is shown in grey at the bottom of \cref{fig:Q-indiv}.
Indeed the spurious features in the individual spectra are lost and a prominent peak emerges near the ground truth.
Although the greatest nominal resolving power occurs near $Q_{\rm opt}\approx q_{\rm max}$, restricting the average near this point is not necessarily advantageous. 
First, the amplitude of the probe ambiguity function decreases with increasing $Q$, weakening the depth response. Therefore, achieving optimal resolving power sacrifices statistical precision due to reduced counts. 
Second, averaging many spectra over a broader range of $Q$ helps suppress spurious features which are unlikely to persist across a diverse set of spectra.

When an iron atom is placed in a 20 nm-thick SrTiO$_3$ crystal oriented along the [001] zone axis in place of a titanium atom 39 Å from the entrance surface, and again $-61$ Å from the focal plane, the coloured spectra overlaid in \cref{fig:Q-indiv} result. As before, the averaged spectrum is provided at the bottom of \cref{fig:Q-indiv} in solid black.
It is apparent that deviations from the Born approximation are present, in particular, spurious features persist across the depth spectra. For instance, the depth spectra of $\vQ=(0.45,0.77),(0.19,-0.96),(-0.70,0.83) {\rm \AA}^{-1}$  contain peaks near $z\sim 50$ Å in common which are represented in the averaged spectrum, albeit with  reduced prominence.

\begin{figure}[!htbp]
    \centering
    \includegraphics[width=1\linewidth]{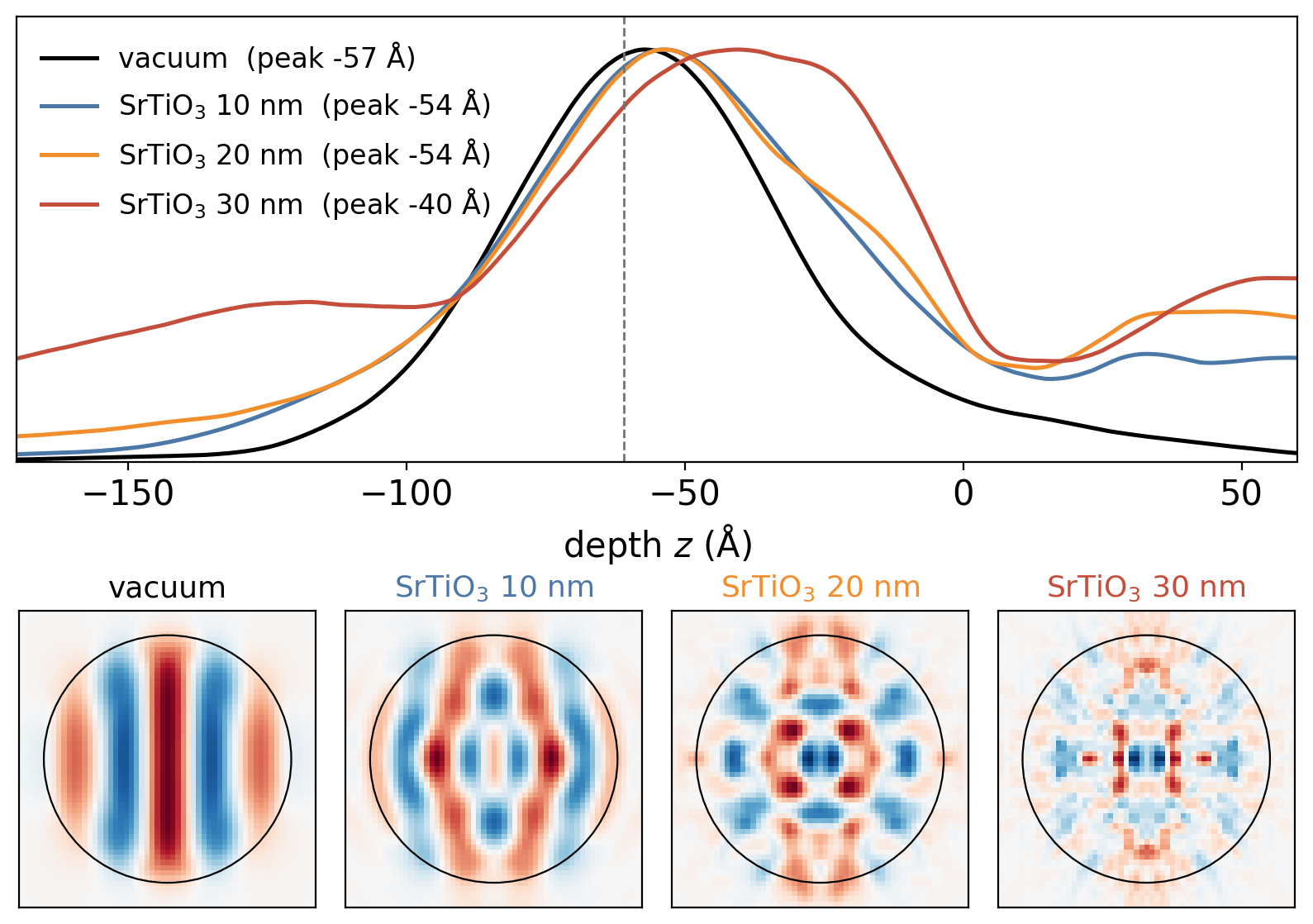}
    \caption{Averaged Fe-L$_{2,3}$ depth spectra in vacuum and [001]-oriented SrTiO\(_3\) of increasing thickness. The dopant lies \(39.1\;{\rm \AA}\) inside the specimen and the defocus is \(-100\;{\rm \AA}\), giving a ground-truth depth relative to focus of \(-60.9\;{\rm \AA}\), marked by the vertical dashed line. The spectra are averaged over \(0.4\leq Q \leq 2.7\;{\rm \AA}^{-1}\) and normalized independently. Increasing specimen thickness strengthens dynamical scattering, progressively distorting the  depth distribution. The lower panels show the corresponding real components of fringe patterns at \(\mathbf Q=(0.77,0)\;{\rm \AA}^{-1}\), with circles marking the bright field aperture. The clean vacuum fringes become increasingly distorted with thickness, culminating in a spurious depth maximum for the thickest specimen. The convergence semi-angle is 30 mrad and the accelerating voltage is 300 keV.
}
    \label{fig:on-zone}
\end{figure}

Unfortunately, while averaging in this instance has extended the validity of the Born approximation to greater thicknesses, averaging is not perpetually immune to channelling artefacts.  
The impact of channelling on the averaged depth spectrum is shown in \cref{fig:on-zone} by progressively increasing the SrTiO$_3$ crystal thickness. The vacuum reference (black) and 20 nm-thick crystal (orange) from \cref{fig:Q-indiv} are reproduced alongside 10 nm (blue) and 30 nm (red) case studies, together with representative fringe patterns at $\vQ=(0.77,0)$\AA$^{-1}$ to provide an additional qualitative measure of deviation from the Born approximation.
% As previously observed, the 20 nm spectrum develops a spurious peak near $50{\rm \AA }$ which is also present with less prominence at 10 nm. 
% Consequently, the fringes at 20 nm are substantially more distorted than at 10 nm. 
% By 30 nm, fringe contrast is entirely lost, and the depth spectrum's dominant peak is displaced from the ground truth, resulting in a biased depth estimate.
The fringes at 20 nm are substantially more distorted than at 10 nm which is evident also in the depth spectra of the 20 nm depth spectrum which develops a stronger spurious peak near $50\;\mathrm{\AA}$, compared with at 10 nm. At 30 nm, the dominant peak is additionally displaced from the ground truth, resulting in a biased depth estimate.

% Fortunately, the interpretation developed under the Born approximation does not rely on the specimen being oriented along a zone axis, where channelling is strongest. 

\begin{figure}[!htbp]
    \centering
    \includegraphics[width=1\linewidth]{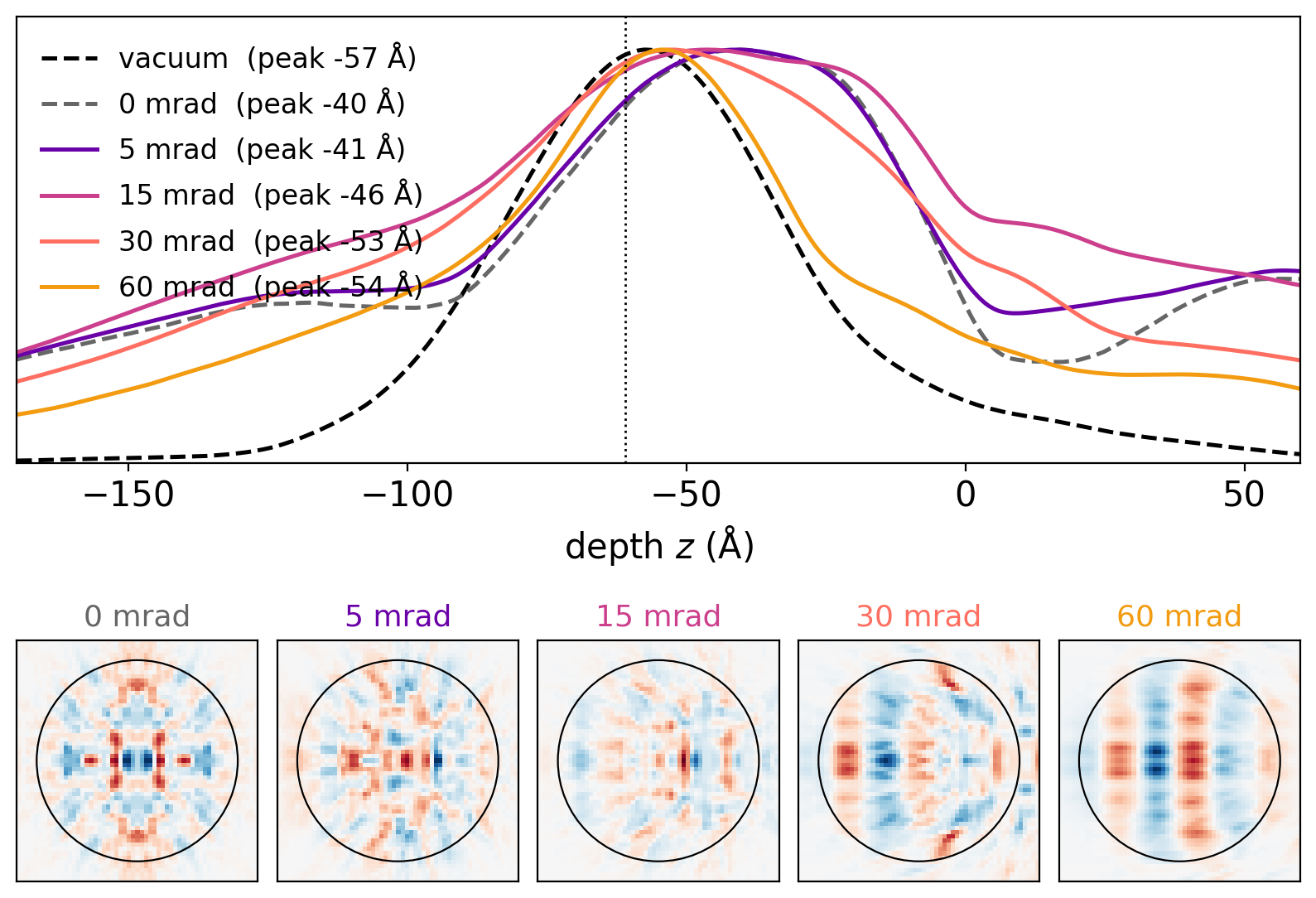}
    \caption{Averaged Fe-L$_{2,3}$ depth spectra from a dopant in vacuum and in \(30\mathrm{nm}\) of SrTiO\(_3\) tilted away from the [001] zone axis. The dopant lies \(39.1\;{\rm \AA}\) inside the specimen with a defocus of \(-100\;{\rm \AA}\), giving a ground-truth depth relative to focus of \(-60.9\;{\rm \AA}\), marked by the dotted line. The vacuum and on-zone references are dashed. Each spectrum is averaged over \(0.4\leq Q \leq 2.7\;{\rm \AA}^{-1}\) and normalized independently. The lower panels show the corresponding real components of fringe patterns at \(\mathbf Q=(0.77,0)\;{\rm \AA}^{-1}\), with the bright field aperture marked by the circle. Tilting disrupts channelling and progressively restores the ordered fringe pattern and depth spectra expected in the kinematical limit. The convergence semi-angle is \(30\mathrm{ mrad}\) and the accelerating voltage is \(300\;\mathrm{keV}\).}
    \label{fig:off-zone}
\end{figure}

Fortunately, zone-axis alignment is not required, allowing channelling to be partly mitigated by tilting the specimen. This is demonstrated in \cref{fig:off-zone} for a 30 nm-thick SrTiO$_3$ specimen at tilts of 5, 15, 30, and 60 mrad away from the [001] zone axis.
The vacuum and on-zone spectra  (0 mrad tilt) are retained as references. 
Tilts of 5 and 15 mrad produce little improvement in depth accuracy, with fringe contrast remaining largely absent. 
At 30 mrad, fringe contrast begins to re-emerge, accompanied by a more substantial shift of the spectral peak towards the ground truth.
 At 60 mrad, prominent fringes are restored and the inferred depth lies well within 1 nm of the ground truth.
% Thus acquisition away from major zone axes indeed further extends the validity of the Born approximation. 

\section{Noise and Bayesian Inference}

\begin{figure}
    \centering
    \includegraphics[width=1\linewidth]{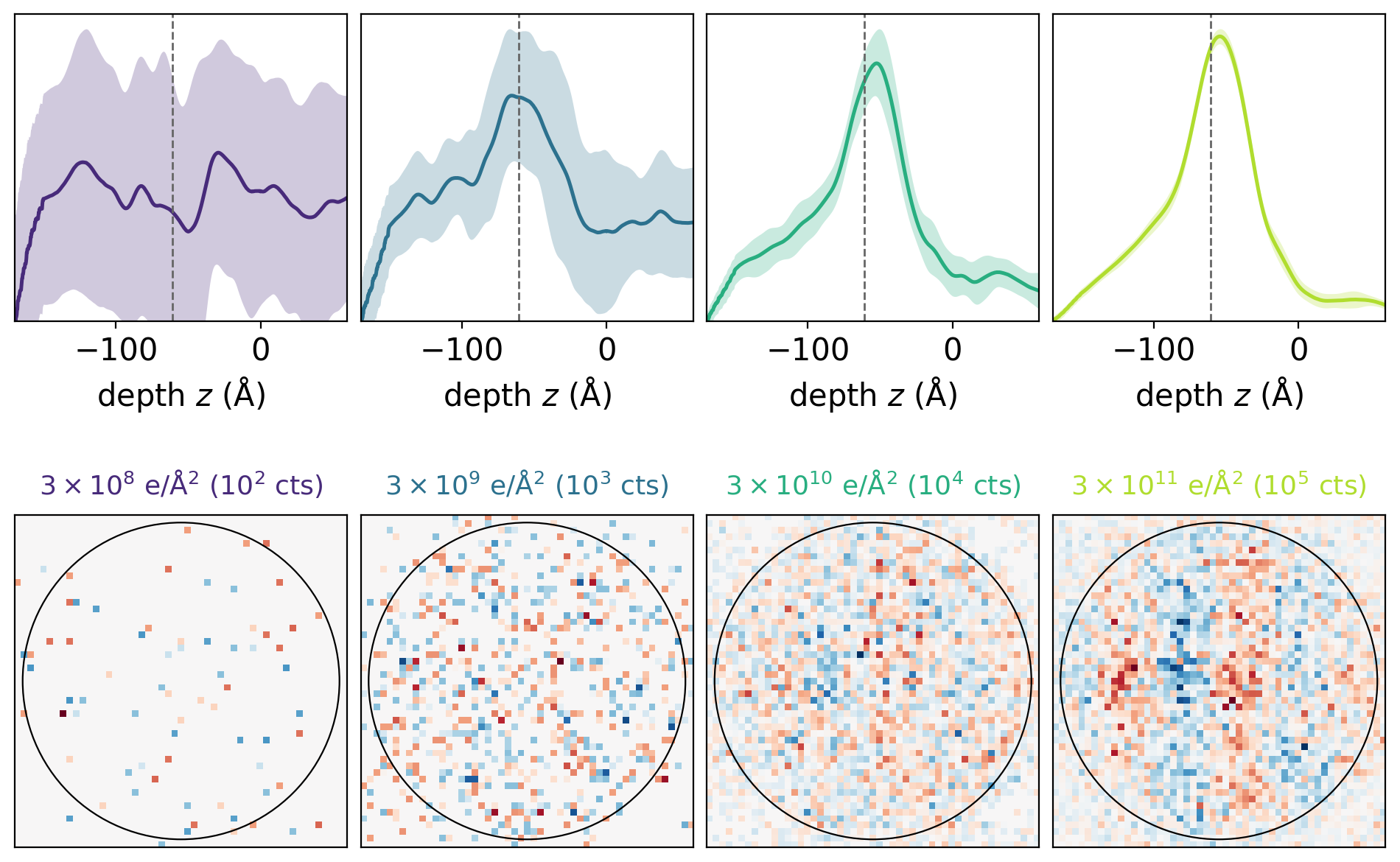}
    \caption{
Fe-L$_{2,3}$ depth spectrum for a dopant in \(30\mathrm{nm}\) of SrTiO\(_3\), tilted \(60\mathrm{ mrad}\) away from the [001] zone axis, and averaged over \(0.4\leq Q \leq 2.7\mathrm{\AA^{-1}}\), constructed from noisy datasets at various doses. 
The dopant lies \(39.1\;{\rm \AA}\) inside the specimen with a defocus of \(-100\;{\rm \AA}\), giving a ground-truth depth relative to focus of \(-60.9\;{\rm \AA}\), marked by the dashed lines. Solid curves show the mean spectrum and shaded regions show \(\pm1\sigma\) across 20 independent Poisson realizations. 
The lower panels show a noisy scan-Fourier pattern at \(\mathbf Q=(0.77,0)\mathrm{\AA^{-1}}\) for each dose, where circles mark the bright field aperture.
The depth peak is obscured by shot noise at $3\times10^{8}\;\mathrm{e/{\rm \AA}^2}$, emerges at $3\times10^{9}$, and sharpens with tight uncertainty at $3\times10^{10}$ and $3\times10^{11}$.
The convergence semi-angle is \(30\mathrm{ mrad}\), the accelerating voltage is \(300\;\mathrm{keV}\).}
    \label{fig:noise}
\end{figure}

Since core-loss scattering is exceedingly rare, populating most bins of a 4D-STEM dataset requires enormous electron doses. Some examples of depth spectra constructed from noisy simulations of the 30 nm-thick SrTiO$_3$ crystal tilted by 60 mrad are shown in \cref{fig:noise}. 
For each of 20 noise realizations, the depth spectrum is averaged across $0.4\leq Q \leq 2.7\mathrm{\AA^{-1}}$. The solid curves indicate the average of the 20 realizations, with the shaded regions representing the corresponding one-standard-deviation spread.
% The effect of each dose is demonstrated on fringes at $Q = (0.77,0)$\AA$^{-1}$, with total doses ranging from $3\times10^{8}$ to $3\times10^{11}$e/\AA$^2$, corresponding to expected core-loss counts across the entire 4D-STEM dataset of $\sim 10^2$ to $\sim10^5$.
Although the fringes remain obscured by noise except at the highest dose of $3\times10^{11}$ e/\AA$^2$, the depth spectrum still contains a distinct peak at $3\times10^{10}$ e/\AA$^2$. At $3\times10^{9}$ e/\AA$^2$, a distinct peak is no longer clearly discernable, and  at $3\times10^{8}$ e/\AA$^2$ no peak remains. 
Despite sacrificing depth resolution to average multiple spectra, the doses required for a well-resolved spectrum present a significant practical barrier.\footnote{The severity may be overstated  since the prominent white lines of the Fe-L$_{2,3}$ edge are not modelled and an effective 1 eV window at 10 eV is assumed by virtue of simulating only 10 eV ionizations above threshold. While this could plausibly reduce the required dose by an order of magnitude or so, still, the doses necessary to recover a well-resolved depth spectrum remain prohibitive for most material systems.}

    % Dose dependence of the  Fe-L$_{2,3}$ depth spectrum for a dopant in \(30\mathrm{nm}\) of SrTiO\(_3\), tilted \(60\mathrm{ mrad}\) away from the [001] zone axis. The dopant lies \(39.1\;{\rm \AA}\) inside the specimen with a defocus of \(-100\;{\rm \AA}\), giving a ground-truth depth relative to focus of \(-60.9\;{\rm \AA}\), marked by the dashed lines. Solid curves show the mean spectrum and shaded regions show \(\pm1\sigma\) across 20 independent Poisson realizations. The total expected core-loss counts are given in parentheses beside each incident dose. At \(3\times10^8\mathrm{e/\AA^2}\), shot noise obscures the depth peak and produces substantial realization-to-realization variation. A peak begins to emerge at \(3\times10^9\mathrm{e/\AA^2}\), while the spectra at \(3\times10^{10}\) and \(3\times10^{11}\mathrm{e/\AA^2}\) are progressively better localized with tightly bounded uncertainty. The lower panels show one noisy scan-Fourier pattern at \(\mathbf Q=(0.77,0)\mathrm{\AA^{-1}}\) for each dose, scaled independently; circles mark the bright field aperture. The convergence semi-angle is \(30\mathrm{ mrad}\), the accelerating voltage is \(300\;\mathrm{keV}\), and spectra are accumulated over \(0.4\leq Q \leq 2.7\mathrm{\AA^{-1}}\).

% Still, the doses required to see a clear depth spectrum remains prohibitive for the many material systems. 
Instead, we infer the dopant's depth using Bayesian inference as a basis for statistical detection.
This approach will further alleviate the demanding dose requirements, rigorously account for background contributions, and substantially improve the achievable resolving power and depth resolution.
Bayesian inference is predicated on Bayes' Theorem 
\begin{equation}
    p(\theta\mid \mathcal{D},M) = \frac{\mathcal{L}(\mathcal{D}\mid \theta,M) \pi(\theta)}{\mathcal{Z}}
    \label{eq:bayesthm}
\end{equation}
where $p(\theta\mid \mathcal{D},M)$ denotes the $a$-$posteriori$ distribution of the quantity to be inferred, $\theta$, $\pi(\theta)$ provides the capacity to systematically incorporate $a$-$priori$ assumptions about the distribution of $\theta$, and $\mathcal{Z}$ serves as a normalization ensuring $p(\theta\mid \mathcal{D},M)$ is a proper density. The likelihood, $\mathcal{L}$, couples the $a$-$priori$ and $a$-$posteriori$ distributions based on observed data, $\mathcal{D}$. Given a forward model, $M$, which predicts the expected counts per bin as a function of $\theta$, and a noise process responsible for the noisy measurements, the likelihood quantifies how well the observed data is explained by a given set of model parameters. 
For our purposes, both the dopant’s spatial coordinates and a nuisance parameter, $\bar{N}$, the expected core-loss counts over all scattering angles, are to be inferred, hence, $\theta = (x, y, z, \bar{N})$.
Unless otherwise stated, all Bayesian inference examples that follow will assume uninformative priors for the dopant’s position. Laterally, the prior is uniform over a $5 \times 5 \;{\rm \AA^2}$ window centred on the dopant column, and in depth it is uniform from -100 to -20 Å from the focal plane. The expected core-loss counts $\bar N$ are assigned a log-uniform prior, so that $\log_{10} \bar N \sim \mathcal{U}(\log_{10} \bar N_{\min}, \log_{10} \bar N_{\max})$, which leaves the inference agnostic to the order of magnitude of the signal. The bounds $(\bar N_{\min}, \bar N_{\max})$ are set to $(10, 150)$, $(100, 1000)$ and $(500, 1500)$ for expected core-loss counts $\bar N=100,500$ and $1000$, respectively.

Counts in distinct scan, detector and energy bins are modelled as statistically independent Poisson variables 
\begin{align}
    \mathcal{D}(\vR,\vq, E) \sim  {\rm Poisson}( \mu(\vR,\vq,E;\theta)).
\end{align}
However, for simplicity, the energy resolution will be eliminated at inference time by summing the mean counts per bin,
\begin{align}
    \mu(\vR,\vq;\theta) &= \sum_{E \in {\rm edge}} \mu(\vR,\vq, E;\theta),
\end{align}
and the 4D-STEM data, 
\begin{align}
    \mathcal{D}(\vR,\vq) =  \sum_{E \in {\rm edge}}  \mathcal{D}(\vR,\vq, E).
\end{align}
Therefore the likelihood is given by
\begin{align}
p(\mathcal{D}|\theta) = \prod_{\vR,\vq\in{\rm BF}} \frac{\mu(\vR,\vq;\theta)^{\mathcal{D}(\vR,\vq)} e^{-\mu(\vR,\vq;\theta)}}{\mathcal{D}(\vR,\vq)!},
\end{align}
and log-likelihood by
\begin{align}
\log p(\mathcal{D}|\theta) = \sum_{\vR,\vq\in{\rm BF}} \Big[ \mathcal{D}(\vR,\vq)\log \mu(\vR,\vq;\theta) - \mu(\vR,\vq;\theta) \Big],
\label{eq:poisson_likelihood}
\end{align}
where the constant, $-\sum \log \mathcal{D}(\vR,\vq)!$ does not depend on $\theta$ and has been dropped.

In the following subsections, we develop the forward model, $M$, which comprises separate models for the background and the core-loss diffraction patterns.

\subsection{Background Modelling}

Two distinct aspects of background modelling, which should not be conflated, must be addressed. First, in practice, an experimental dataset contains a background component that must be estimated by fitting an empirical model before edge integration. Second, for simulated demonstrations, a model for the momentum-resolved  background is required to generate synthetic 4D-STEM datasets. We will assume that during inference the background is known exactly from having performed empirical fits.
% Each of these aspects is treated in turn separately below.

\subsubsection{Pre-Edge Fitting in Practice}

In practice, the background in a 4D-STEM dataset can be isolated in the pre-edge from the core-loss signal. Hence, before integrating over an edge window, $B(\vR,\vq, E)$ is estimated by assuming the background spectrum can be modelled with a power-law \citep{egerton1975inelastic}, 
\begin{align}
    B(\vR,\vq,E) = A(\vR,\vq) E^{-\gamma}
    \label{eq:BAE}
\end{align}
where $A(\vR,\vq)$ captures the background's dependence on probe position and detector reciprocal coordinate.
As such, a pre-edge window is used to fit a power-law which is extrapolated beneath the core-loss edge and subtracted from the observed counts. This procedure, while commonplace and straightforward, has garnered criticism  \citep{egerton2002improved,su1993background,pun1984optimized,pun1985weighted,unser1987optimal,liu1987influence}. 
In particular, subtracting a fitted background from the observed counts does not recover a Poisson-distributed signal. A likelihood constructed from the background-subtracted remainder must therefore account for the altered noise distribution, particularly in the low-count regime where Poisson noise is poorly approximated as additive.
As done by \cite{verbeeck2004model}, inference will be performed on the background-retained dataset to ensure a Poisson likelihood remains valid. This treats the noise properly at all count levels, including the low-count regime that is most relevant for core-loss signals.

Following \citet{deimetry2026threedimensionaldopantdistributionsingle} and \citet{unser1987optimal}, we assume the same background power-law decay constant, $\gamma$, across the entire dataset and estimate it by fitting the momentum-integrated pre-edge spectrum using least-squares. Given $\gamma$, the Poisson maximum likelihood estimate (MLE) of the background amplitude of each bin is given by  \citep{deimetry2026threedimensionaldopantdistributionsingle}
\begin{align}
    A(\vR,\vq)=\frac{\sum_{E \in \mathrm{pre-edge}} \mathcal{D}(\vR,\vq,E)}{\sum_{E \in \mathrm{pre-edge}} E^{-\gamma}}.
\end{align}
That is, the power-law amplitude is given by the ratio of the observed counts to the unit-amplitude background power-law summed over the pre-edge window. 
Although the uncertainty in the MLE should, in principle, be propagated through the Bayesian inference pipeline, doing so would require jointly inferring every background element alongside the signal, for instance, using priors informed by the background MLE and its uncertainty. The computational cost of such an approach would be prohibitive, so the synthetic background is assumed known and to be error free.

\subsubsection{Modelling Momentum-Resolved Background Intensities for Simulation}

Quantifying the degradation of dopant localization precision in simulation due to background requires a momentum-resolved background model for generating synthetic edge-integrated 4D-STEM datasets, and a definition of the SBR applicable to non-uniform scattering distributions.

\begin{figure*}
    \centering
    \includegraphics[width=1\linewidth]{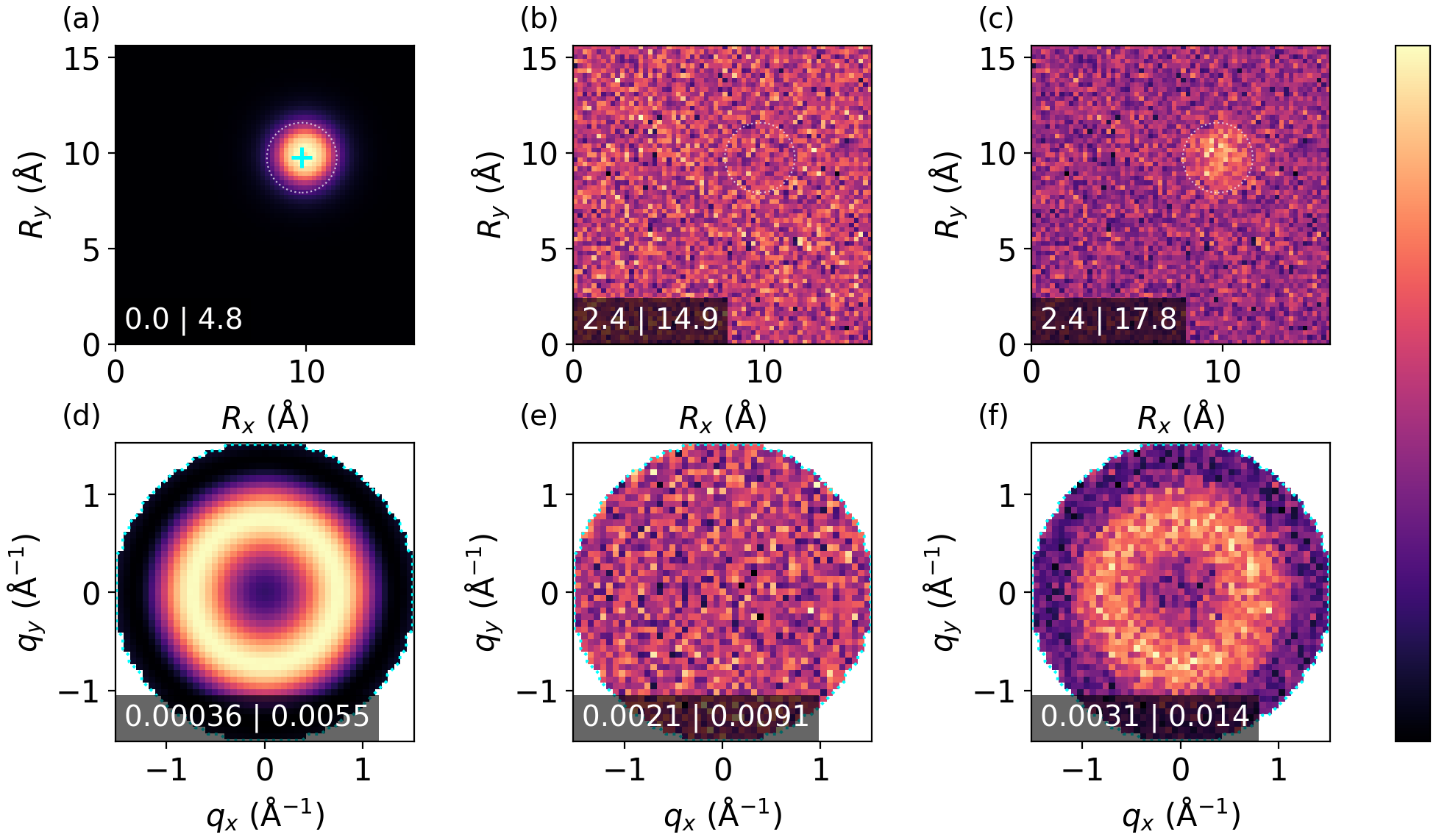}
    \caption{Construction of the synthetic background for $\bar N=1000$ expected core-loss counts and $\mathrm{SBR}=0.5$. (a) Background-free STEM-EELS image of a single Fe atom in vacuum, integrated over bright field. (b) Spatial distribution of the expected background counts, generated according to \cref{eq:Btotal,eq:normal-draw}. (c) STEM-EELS image and background superimposed. An example (d) core-loss diffraction pattern at the probe position marked by the cross in (a) located at (9.8, 9.8) Å. (e) Synthetic momentum-resolved background, as specified in \cref{eq:Bmomres}. (f) Combined core-loss and background diffraction pattern. All panels represent expected counts prior to Poisson noise realization. Each panel is independently scaled to the colour bar on the right, with its minimum and maximum intensities indicated in the lower-left corner as $\langle\mathrm{min}\rangle\mid\langle\mathrm{max}\rangle$. }
    \label{fig:bkg-demo}
\end{figure*}

Shown in \cref{fig:bkg-demo}(a) is an idealized core-loss STEM-EELS image acquired with a detector spanning the bright field. The field of view appears empty except for a single disk centred laterally on the dopant with radius proportional to the dopant's distance from the focal plane  \citep{deimetry2026threedimensionaldopantdistributionsingle}. Within this disk the intensity is approximately uniform so the SBR can be defined as  
\begin{equation}
    {\rm SBR} = \frac{\bar N_{\rm BF}}{\bar  N_{B_{\rm disk}}}
\end{equation}
where $\bar N_{\rm BF}$ is the mean core-loss counts across the entire dataset within the detector (here, the bright field), and $\bar  N_{B_{\rm disk}}$ is the mean background counts within the same detector range, summed over probe positions within the disk only. For a probe of convergence semi-angle, $\alpha$, the disk has radius $\alpha |z|$, where $z$ is the distance of the dopant from the focal plane, and so occupies a fraction
\begin{align}
    f_{\rm disk} = \frac{\pi \alpha^2 z^2}{L_x L_y}
\end{align}
of the scan area, where $L_x, L_y$ are the dimensions of the scan region. The generating depth and expected core-loss counts are used to set the synthetic background, which is then held fixed during inference. Assuming the background to be uniform over the scan, the total background across the scan region is
\begin{align}
   \bar  N_{B_{\rm total}} = \frac{\bar N_{\rm BF}}{f_{\rm disk} {\rm SBR}}.
    \label{eq:Btotal}
\end{align}
That is, the SBR prescribes the background within the disk which is scaled by $f_{\rm disk}$ to determine the total counts required to uniformly fill the scan field of view.

In practice, the background is expected to vary smoothly with probe position, whereas \cref{eq:Btotal} assumes a constant background across the scan region. Explicitly modelling this spatial variation is beyond the scope of this work. Instead, an intermediate approach is adopted that relaxes the assumption of uniformity without imposing a particular spatial dependence. Specifically, the uniform background level in \cref{eq:Btotal} is taken as the mean of a normal distribution, from which the expected background counts at each probe position are independently drawn
\begin{align}
    B(\vR) \sim \text{Normal}\left(\mu=\frac{\bar  N_{B_{\rm total}}}{n_{\rm probes}},\sigma=0.2\mu \right),
    \label{eq:normal-draw}
\end{align}
where $n_{\rm probes}$ is the number of probe positions.
A standard deviation of 20\% introduces spatial variation in the background while preserving, on average, the prescribed total background. As $  B(\vR)$ is non-negative, negative draws are set to zero.
 \Cref{fig:bkg-demo}(b) illustrates this procedure for SBR=0.5 and $\bar N=1000$. We stress that the background looks noisy because the mean value is drawn from \cref{eq:normal-draw}, not because of shot noise. The STEM-EELS image with this background added is shown in \cref{fig:bkg-demo}(c).

% momentum res in background
Having specified the variation of the background across probe positions, we must also prescribe a distribution across momentum space. 
% This requires some care since the apparent SBR can become artificially dependent on the extent of the detector. In particular, distributing a fixed background over an increasingly large detector would dilute its contribution within the bright-field region, where most of the core-loss signal is concentrated, thereby exaggerating the apparent prominence of the signal. 
Similarly,
\begin{align}
    B(\vR,\vq) \sim  \text{Normal}\left(\mu=\frac{B(\vR)}{n_{\rm BF}},\sigma=0.2\mu \right).
    \label{eq:Bmomres}
\end{align}
where $n_{\rm BF}$ denotes the number of pixels in the bright field, and again, setting negative values to zero. 
% The complete background is then rescaled to the total in \cref{eq:Btotal}.
An example of a synthetic background assuming $\mathrm{SBR}=0.5$ and $\bar N=1000$ is shown in \cref{fig:bkg-demo}(e), which, superimposed onto the diffraction pattern in \cref{fig:bkg-demo}(d), yields the total signal in \cref{fig:bkg-demo}(f).

% An additional example for  $\bar N=100$ core-loss counts at (a) $\mathrm{SBR}=0.1$ and (b) $\mathrm{SBR}=0.5$ is given in \cref{fig:ex-means}. At the lower SBR, the core-loss distribution is largely obscured by the background, whereas at SBR=$0.5$, its characteristic structure becomes clearly discernible within the bright-field region. Note, the latter closely resembles \cref{fig:bkg-demo}(f), albeit with both the core-loss signal and background reduced proportionally to preserve the same SBR at a lower core-loss count.

% \begin{figure}
%     \centering
%     \includegraphics[width=1\linewidth]{inference_20261103/background_sweep.png}
%     \caption{ Influence of the SBR on the visibility of the core-loss signal for $\bar N=100$ expected core-loss counts. Momentum-resolved diffraction patterns showing the combined core-loss and synthetic background intensities at (a) $\mathrm{SBR}=0.1$ and (b) $\mathrm{SBR}=0.5$. Each panel is independently scaled to the colour bar on the right, with its minimum and maximum intensities indicated in the lower-left corner as $\langle\mathrm{min}\rangle\mid\langle\mathrm{max}\rangle$. }
%     \label{fig:ex-means}
% \end{figure}

\subsection{Momentum-Resolved Core-Loss Modelling}

With the background accounted for, all that remains is a model for the core-loss intensity $I(\vR,\vq;\theta)$ after summing over an edge window. 
For optimal fidelity, the multislice algorithm should be used.
However, the Born model is computationally cheaper and does not rely on host-structure parameters. The consequences of using the Born approximation will be revisited in \cref{sec:simstatinf}.

The forward model predicts the core-loss signal only up to an overall scale, which depends on the dose. Therefore, we define the normalized intensity \footnote{The normalization is a discretized form of \cref{eq:Nnorm}.}
\begin{align}
    \bar I(\vR,\vq;\theta)=\frac{I(\vR,\vq;\theta)}{\sum_{\vR,\vq} I(\vR,\vq;\theta)},
    \label{eq:barIdef}
\end{align}
such that $\sum_{\vR,\vq}\bar I=1$.
% and treat the total core-loss counts, $\bar N$, as a nuisance parameter to be inferred alongside the dopant's location.
The intensity integrated over probe position and the detector gives\footnote{Summing incoherently over transitions, the mod square of \cref{eq:apexp1} gives
\begin{equation}
\begin{aligned}
I(\mathbf{R},\mathbf{q};\theta)&=\sum_{fi}\sigma_{fi}^2\int d^2q'\,d^2q''\, e^{-2\pi i(\mathbf{q}''-\mathbf{q}')\cdot\mathbf{x}}\, \\
&\quad \times (\hat t\ast\hat H_{fi})(\mathbf{q}-\mathbf{q}')\, (\hat t\ast\hat H_{fi})^*(\mathbf{q}-\mathbf{q}'') \\
&\quad \times e^{-i\pi\lambda z(q'^2-q''^2)}\, \hat\psi_0(\mathbf{q}')\,\hat\psi_0^*(\mathbf{q}'')\, e^{-2\pi i(\mathbf{q}'-\mathbf{q}'')\cdot\mathbf{R}}.
\end{aligned}
\label{eq:Iexpand}
\end{equation}
Using $\int d^2R\,e^{-2\pi i(\mathbf{q}'-\mathbf{q}'')\cdot\mathbf{R}}=\delta^2(\mathbf{q}'-\mathbf{q}'')$, the $\mathbf{x}$ and $z$ phases cancel and
\begin{equation}
\int d^2R\,I=\sum_{fi}\sigma_{fi}^2\int d^2q'\,|(\hat t\ast\hat H_{fi})(\mathbf{q}-\mathbf{q}')|^2\,|\hat\psi_0(\mathbf{q}')|^2 .
\end{equation}
Restricting $\mathbf{q}$ to the detector and integrating yields \cref{eq:normalization}.
}
\begin{equation}
\int d^2\mathbf{R}\,\int_{\rm det} d^2\mathbf{q}\,I(\mathbf{R},\mathbf{q};\theta) = \sum_{fi}\sigma_{fi}^2\int_{\rm det}d^2\mathbf{q}\int d^2\mathbf{q}' \big|(\hat t\ast\hat H_{fi})(\mathbf{q}-\mathbf{q}')\big|^2\, |\hat\psi_0(\mathbf{q}')|^2 .
\label{eq:normalization}
\end{equation}
which reduces to
\begin{equation}
\int d^2R\,d^2q\,I(\vR,\vq;\theta) =\left(\int d^2k\,|\hat\psi_0(\mathbf{k})|^2\right) \left(\sum_{fi}\sigma_{fi}^2\int d^2q\, \big|(\hat t\ast\hat H_{fi})(\mathbf{q})\big|^2\right).
\label{eq:Nnorm}
\end{equation}
for a detector of infinite extent. 
Note that \cref{eq:normalization,eq:Nnorm}  are independent of the dopant's location and so can be evaluated without constructing the full 4D-STEM dataset. However, \cref{eq:normalization,eq:Nnorm} assume that the integral over probe position runs over all space which is well approximated only when the scan region contains the core-loss disk, such as in \cref{fig:bkg-demo}(a).
For inference, we use a \(15.62\times15.62\,\mathrm{\AA}^2\) scan sampled at \(64\times64\) probe positions, chosen to encompass the largest expected core-loss disk allowed by the depth prior throughout the allowed prior lateral position range.

% Assuming these priors, we choose a $15.62\times15.62$~\AA$^2$ scan which contains the largest geometrical disk allowed by the prior of $6 \; {\rm \AA}$, regardless of where the dopant sits within the  $5 \times 5 \;{\rm \AA^2}$ lateral prior.
% This not only ensures the integral over probe position is well approximated as extending over all space but also ensures no core-loss signal is discarded with the prior information given.

Now, observe that $\mathcal{D}(\vR,\vq)$ causes the first term of \cref{eq:poisson_likelihood} to vanish for empty bins and the second term is given by 
\begin{align}
    \sum_{\vR,\vq\in {\rm det}}\mu(\vR,\vq;\theta)
    &= \sum_{\vR,\vq \in {\rm det}}\left[
       \bar  N\bar I(\vR,\vq;\theta)+B(\vR,\vq)
    \right] \nonumber\\
    &= \bar  N_{\rm det}+\sum_{\vR,\vq \in {\rm det}}B(\vR,\vq).
\end{align}
where $\mathbf q\in{\rm det}$ restricts the sum to reciprocal-space pixels in the detector and $ \bar  N_{\rm det}$ is the expected number of core-loss counts in the detector range which can be calculated independently of the dopant's location by multiplying the ratio of \cref{eq:normalization} to \cref{eq:Nnorm} by $\bar N$, which is inferred. 
Therefore, full dataset normalization can be calculated once independently of where the dopant is located and the core-loss signal forward model only needs to supply predictions for bins with non-zero counts for the likelihood to be evaluated. The intensity of each bin is given by \footnote{Note that the transmission function which was neglected in previous expressions is restored in \cref{eq:infFM} to minimize systematic errors.}
\begin{equation}
\begin{aligned}
\bar I(\vR,\vq;\theta) &= \frac{1}{\sum_{\vR,\vq}I} \sum_{fi}\Bigg| \sum_{\vk\in{\rm BF}} (\hat t \conv (-i \sigma_{fi} )\hat H_{fi})(\vq-\vk)\hat\psi_0(\vk) \\
&\quad \times \exp\Big[ -i\pi\lambda z k^2 -2\pi i(\vq-\vk)\cdot\vx -2\pi i\vk\cdot\vR \Big]\Bigg|^2.
\end{aligned}
\label{eq:infFM}
\end{equation}
This allows the likelihood to be evaluated quickly in the core-loss regime, where sparsity is expected to yield a few counts per diffraction pattern.
To further speed up model evaluation, the sum over states $fi$ can be reduced using the spectral decomposition (SD) specified in \cite{DWYER201568}. This SD decomposition can also incorporate the energy above the ionization threshold quantum number to compress the number of transitions required for edge integration.

\subsection{Simulated Statistical Inference}\label{sec:simstatinf}

% ---------------- inference accuracy vs tilting test ----------------

\begin{figure*}
    \centering
    \includegraphics[width=1\linewidth]{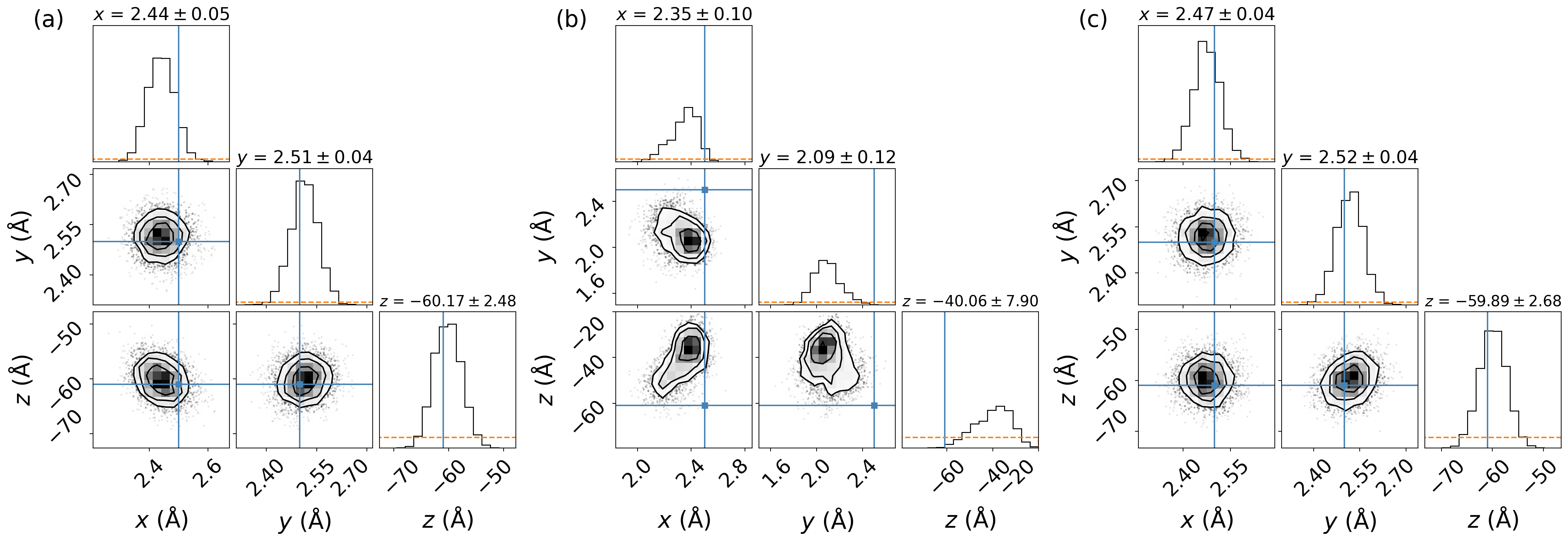}
    \caption{Posterior distributions of dopant 3D location $(x,y,z)$ for (a) dopant isolated in vacuum and  embedded in 30 nm-thick SrTiO$_3$ sample (b) on [001] zone and (c) tilted 60 mrad off-zone. The dopant is located 39 Å from the sample entrance surface. A 30 mrad probe at 300 keV is underfocused into the sample by 100 Å. This places the dopant $61 \;\rm{\AA}$ from the focal plane. Blue lines mark the ground-truth position. Panel titles give the posterior mean $\pm$ standard deviation in angstroms. The expected core-loss count $\bar N$ is a nuisance parameter and is omitted. All cases use $\bar N=1000$ expected core-loss counts and an SBR of 0.5. In vacuum and with 60 mrad tilt, the inference recovers the ground truth. With no tilt, dynamical scattering biases the posterior. Uniform priors were placed on the lateral position, $x, y \sim \mathcal{U}(0, 5)\,\AA$, and on the depth, $z \sim \mathcal{U}(-100, -20)$ Å, while a log-uniform prior was placed on the core-loss counts, $\log_{10} \bar N \sim \mathcal{U}(\log_{10} 500, \log_{10} 1500)$. Priors indicated in marginals in orange. Nested sampling was run with 500 live points and terminated with stopping criterion $\Delta\log \mathcal{Z} < 0.05$. Marginalized posterior titles specify mean and standard deviation. Lateral coordinates marginals share a common vertical scale across sub-figures, separately from z-coordinate.}
    \label{fig:tilting_inference}
\end{figure*}

Having fully specified the details of the inference, we revisit the validity of the Born approximation as a forward model in the presence of dynamical scattering. 
To isolate the influence of dynamical scattering on the accuracy of the inference, \cref{fig:tilting_inference} presents the posterior for (a) an isolated iron atom in vacuum, serving as an idealized reference, and for the same dopant embedded in a 30 nm-thick SrTiO$_3$ matrix, imaged (b) on zone and (c) 60 mrad off zone, with $\bar N=1000$ and SBR=0.5.
In vacuum, the ground-truth position of the atom is faithfully recovered both laterally and along the optic axis with sub-angstrom and sub-nanometre precision, respectively. As anticipated from the studying depth spectra, embedding the dopant in 30 nm-thick SrTiO$_3$ biases the depth posterior, which is largely mitigated by tilting 60 mrad away from the [001] zone axis.

% ---------------- Investigating counts and SBR on inference accuracy at 60 mrad offset  ----------------

\begin{figure*}
    \centering
    \includegraphics[width=1\linewidth]{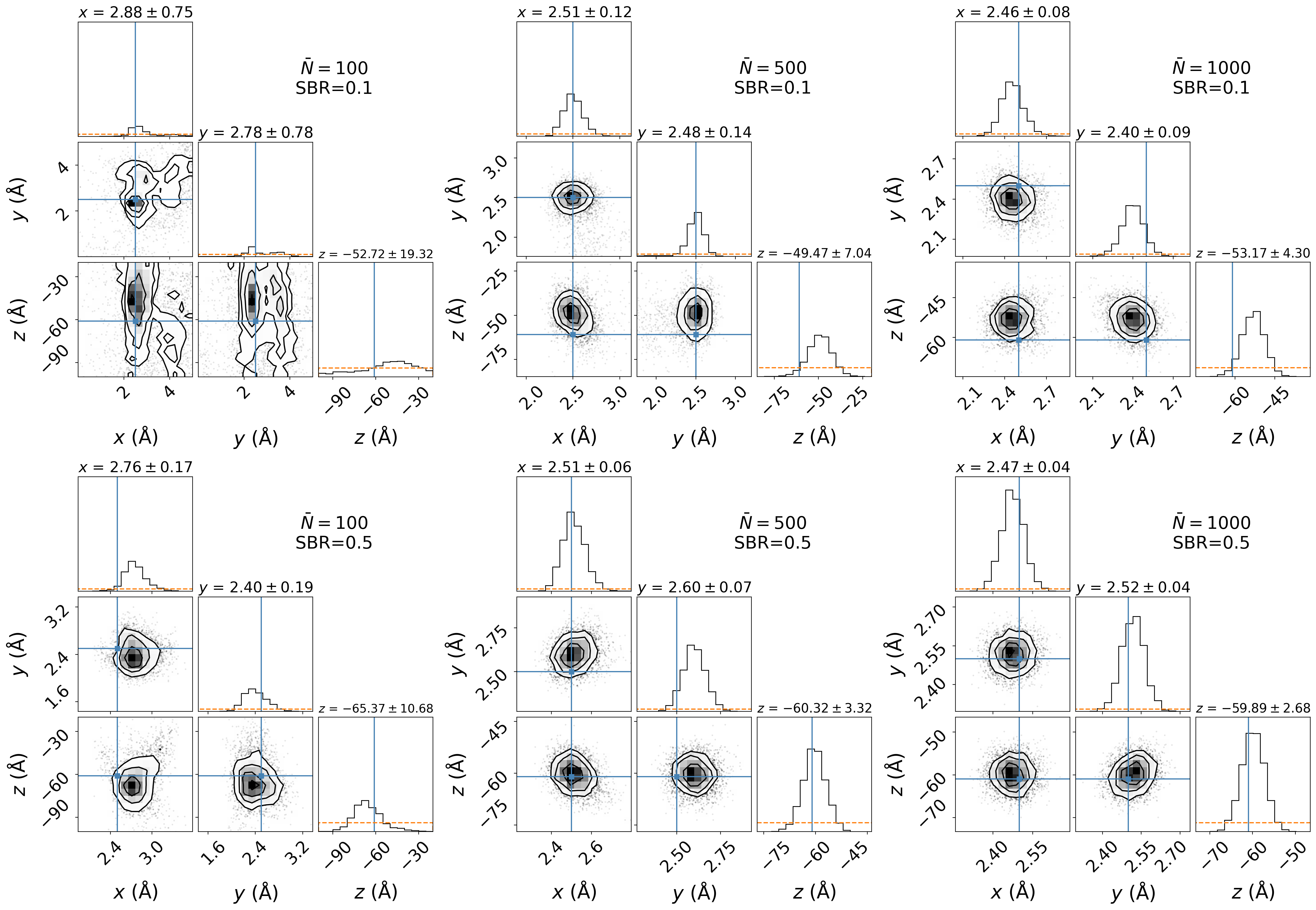}
    \caption{
   Posterior distributions of the dopant position $(x, y, z)$ for varying counts (columns: 100, 500, and 1000) and SBR (rows: 0.1 and 0.5), with 30 mrad probe at 300 keV underfocused into the sample by 100 Å. This places the dopant $61 \;\rm{\AA}$ from the focal plane. Blue lines mark the ground-truth position. Panel titles give the posterior mean $\pm$ standard deviation in angstroms. The expected core-loss count $\bar N$ is a nuisance parameter and is omitted. Uniform priors were placed on the lateral position, $x, y \sim \mathcal{U}(0, 5)\,\AA$, and on the depth, $z \sim \mathcal{U}(-100, -20)$ Å. A log-uniform prior was placed on the expected core-loss counts $\bar N$, such that $\log_{10} \bar N \sim \mathcal{U}(\log_{10} \bar N_{\min}, \log_{10} \bar N_{\max})$, with $(\bar N_{\min}, \bar N_{\max}) = (10, 150)$, $(100, 1000)$ and $(500, 1500)$ for generating expected counts $\bar N=100$, $500$ and $1000$, respectively. Priors indicated in marginals in orange. Nested sampling was run with 500 live points and terminated with stopping criterion $\Delta\log \mathcal{Z} < 0.05$. Marginalized posterior titles specify mean and standard deviation. Lateral coordinates marginals share a common vertical scale across sub-figures, separately from z-coordinate.}
    \label{fig:inference}
\end{figure*}

% In a realistic configuration that is adequately described within the Born approximation, the precision with which the dopant location can be inferred depends only on the SBR and the number of core-loss counts. 
Posterior distributions for the 60 mrad-tilted 30 nm SrTiO$_3$ sample across a range of core-loss counts and SBR values, are shown in \cref{fig:inference}. Since the lateral coordinates behave similarly, the lateral coordinates' marginals share a common vertical scale across corner plots, separately from the z-coordinate.
% We examine the lateral coordinates, which behave similarly, separately from the z-coordinate. As such, the lateral coordinates' marginals share a common vertical scale across corner plots separately from the z-coordinate. 

Note, as core-loss counts and the SBR increase, the posterior distributions increasingly diverge from the uniform prior, indicating that the data becomes increasingly informative.
\Cref{tab:counts_transitions} lists the fold reduction in posterior uncertainty obtained by increasing the expected core-loss counts at fixed SBR. Because shot noise alone predicts an uncertainty reduction of $\sqrt{\bar N_{\rm new}/ \bar N_{\rm old}}$, the parenthetical values normalize the observed improvement by this factor, isolating gains beyond shot noise alone. 
Similarly, \cref{tab:sbr_transitions} lists the fold reduction in posterior uncertainty obtained by raising the SBR from 0.1 to 0.5 at fixed core-loss counts. Because no analogous scaling exists for the SBR, the raw improvement factors without a shot-noise baseline are reported.
% explain
When SBR=0.1 and $\bar N=100$, increasing either SBR or $\bar N$ significantly improves the precision of the lateral coordinates. However, additional improvements to SBR and $\bar N$ yield diminishing returns. Alternatively, the z-coordinate continues to improve steadily with increasing SBR and $\bar N$. This is to be expected since the lateral coordinates only require identifying the probe positions which yield deviations from the expected background distribution while depth requires interpreting the specific momentum redistribution due to the core-loss interaction.

% At SBR = 0.1, the lateral coordinates improve substantially when the counts are increased from 100 to 500, but diminishes sharply from 500 to 1000. At the higher SBR of 0.5, increasing the counts across the full 100 to 1000 range has a weaker effect compared with the lower SBR. 

% The reductions depend on the coordinate, SBR and count level, as quantified in \cref{tab:counts_transitions}.

\begin{table}[ht]
\centering
\caption{Fold reduction in posterior uncertainty in \cref{fig:inference}, $\sigma_{\mathrm{old}}/\sigma_{\mathrm{new}}$,
obtained by increasing the expected core-loss counts at fixed SBR. Values in parentheses are
normalized by the shot-noise expectation $\sqrt{\bar N_{\mathrm{new}}/\bar N_{\mathrm{old}}}$
($2.24$ for $100\rightarrow500$ and $1.41$ for $500\rightarrow1000$ counts).}
\label{tab:counts_transitions}
\begin{tabular}{llccc}
\toprule
SBR & $\bar N$ transition & $x$ & $y$ & $z$ \\
\midrule
\multirow{2}{*}{0.1} & $100 \rightarrow 500$ & 6.14 (2.74) & 5.43 (2.43) & 2.75 (1.23) \\
 & $500 \rightarrow 1000$& 1.49 (1.05) & 1.69 (1.20) & 1.64 (1.16) \\
\midrule
\multirow{2}{*}{0.5} & $100 \rightarrow 500$ & 2.91 (1.30) & 2.97 (1.33) & 3.22 (1.44) \\
 & $500 \rightarrow 1000$ & 1.32 (0.94) & 1.47 (1.04) & 1.24 (0.88) \\
\bottomrule
\end{tabular}
\end{table}

\begin{table}[ht]
\centering
\caption{Fold reduction in posterior uncertainty in \cref{fig:inference}, $\sigma_{\mathrm{old}}/\sigma_{\mathrm{new}}$,
obtained by increasing the SBR from 0.1 to 0.5 at fixed expected core-loss counts.}
\label{tab:sbr_transitions}
\begin{tabular}{lccc}
\toprule
$\bar N$ & $x$ & $y$ & $z$ \\
\midrule
100 & 4.52 & 4.05 & 1.81 \\
500 & 2.14 & 2.22 & 2.12 \\
1000 & 1.91 & 1.92 & 1.61 \\
\bottomrule
\end{tabular}
\end{table}

% ---------------- Additional benifit ---------------- 

An additional benefit of a statistical approach is that the number of dopants can be examined systematically using so-called Bayes factors. Observe the normalization, $\mathcal{Z}$, in \cref{eq:bayesthm} is given by 
\begin{align}
    \mathcal{Z} = p(\mathcal{D}\mid M).
\end{align}
That is, the normalization is given by the probability of obtaining the observed data given the underlying model $M$. Hence, the evidence can be used to discriminate between models according to how well each is supported by the data. In our case, each model corresponds to a different number of dopants, denoted with a subscript, e.g. $M_1$ and $\mathcal{Z}_1$ correspond to a single dopant model and the corresponding evidence. This can be used to test the resolving power of the technique as was done using the Rayleigh criterion in \cref{eq:zrayleigh}. For simplicity, multiple atoms are simulated individually by neglecting scattering from the elastic potential of all other atoms. The momentum-resolved signal from all atoms is approximated as the incoherent sum of intensities from each isolated atom, an approximation that has been shown to be accurate for core-loss scattering from a few atoms  \citep{deimetry2026bayesian}.

\begin{figure}
    \centering
    \includegraphics[width=1\linewidth]{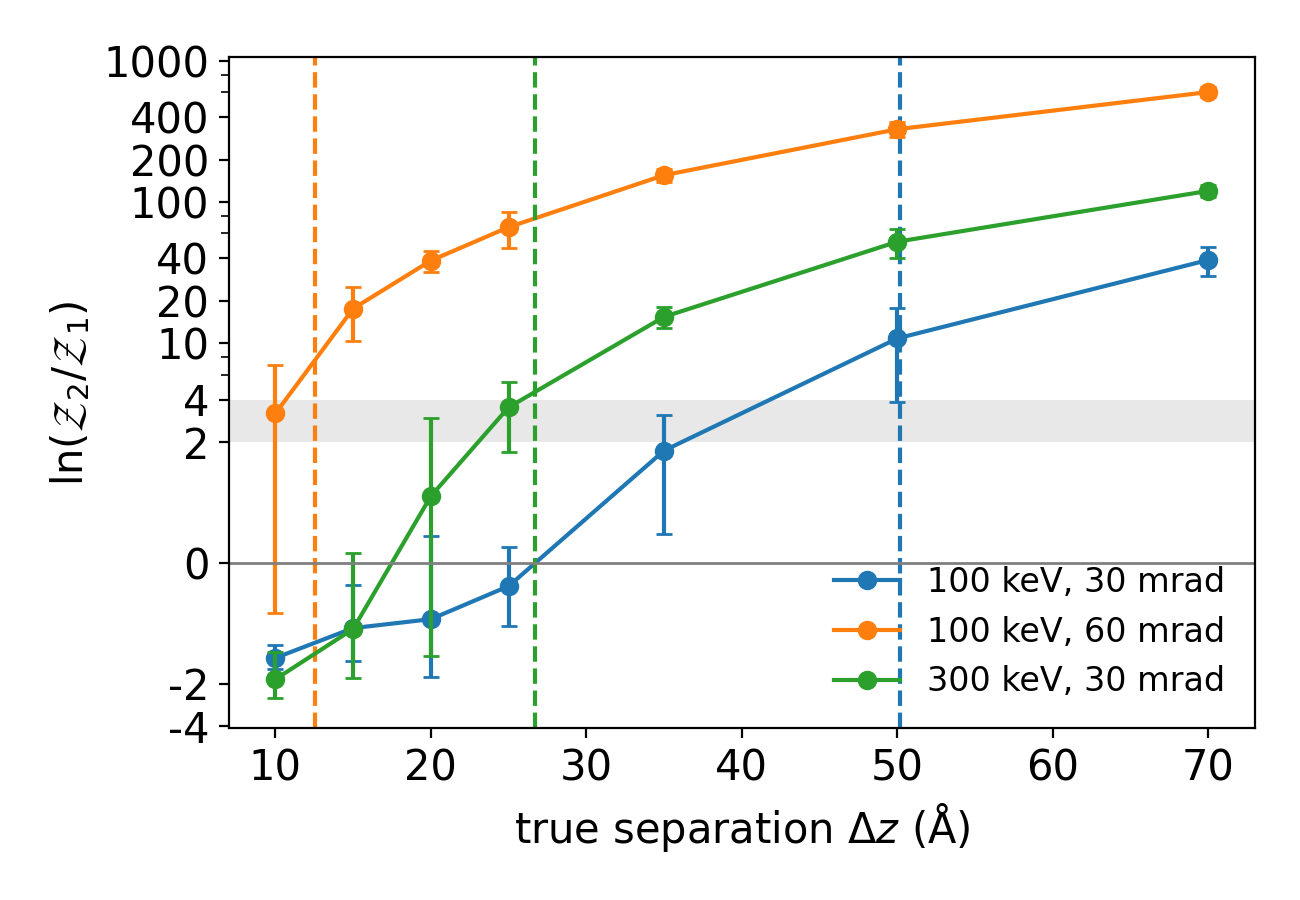}
\caption{Log Bayes factors, $\log(\mathcal{Z}_2/\mathcal{Z}_1)$, comparing the evidence for single-dopant model ($\mathcal{Z}_1$) with the two-dopant model ($\mathcal{Z}_2$), with uncertainty bars showing $\pm 1$ standard deviation calculated from 5 realizations. Both models are evaluated on datasets generated with $\bar N=1000$ expected total core-loss counts containing two dopants on the same column, with increasing separation along the optic axis. Bayes factors are plotted as a function of separation for several accelerating voltages and probe apertures. A conservative bound on the resolving power is set by the Rayleigh criterion in \cref{eq:zrayleigh}, denoted with vertical bars. In the two-dopant model the dopants lie at depths $z_1$ and $z_2 = z_1 + \Delta z$ with $z_1 \sim \mathcal{U}(-50, 50)\,\AA$ and $\Delta z \sim \mathcal{U}(0, 100)\,\AA$, so that the ordering $z_1 < z_2$ is built into the parameterization and the label degeneracy is removed. Both models use a log-uniform prior on the expected total core-loss counts, $\log_{10} \bar N \sim \mathcal{U}(\log_{10} 1, \log_{10} 5000)$, and no background is included. Evidences computed by nested sampling as implemented in Dynesty  \citep{sergey_koposov_2026_21417233,2020MNRAS.493.3132S} used 500 live points and stopping criterion $\Delta\ln\mathcal{Z} < 0.1$. Detector is confined to bright field.}
    \label{fig:bayes}
\end{figure}

The resolving power is tested with two atoms in vacuum at the same lateral position but different depths. The Bayes factor for models assuming a single and two dopants is shown in \cref{fig:bayes}, plotted as a function of dopant separation for several combinations of accelerating voltage and probe aperture semi-angles averaged over several noise realizations. The Rayleigh limit in \cref{eq:zrayleigh} is indicated with vertical bars for comparison. 
While there are several interpretations of Bayes factors for what is considered statistically significant  \citep{jeffreys1998theory,kass1995bayes}, typically a log Bayes factor of around $\sim$2-4 is commonly taken as the threshold for preferring the two-dopant model. 
This range is indicated by the horizontal shaded region in \cref{fig:bayes}. For the resolving power of the Bayesian inference to agree with that predicted by the Rayleigh criterion, the vertical bar would need to intersect the Bayes factor curves within this region. In every instance, it does not. 
Instead, the intersections occur in the range $\ln \mathcal{Z}_2/\mathcal{Z}_1=5$ to $10$, corresponding to very strong evidence.
This outcome is unsurprising, as the present statistical approach exploits information encoded in the full forward model rather than the geometry of the resulting spectrum alone. 
Nevertheless, the Rayleigh criterion correctly anticipates the order that the curves are stacked. 
% Nevertheless, the Rayleigh criterion remains a useful gauge of general trends. 
% First, the magnitude and decay of the uncertainties with increasing separation follow the ordering it predicts, for instance, at 100 keV and 60 mrad (orange curve), which has the smallest Rayleigh limit, exhibits the most rapidly decaying uncertainties relative to the mean value. Second, the curves are stacked in the order that the Rayleigh criterion anticipates.

\section{Conclusion}

We have demonstrated that, within the Born approximation, Fourier transformation of core-loss 4D-STEM over probe position reveals fringes whose frequency encodes dopant depth. The resulting depth spectrum is governed by the characteristics of the ambiguity functions of the probe and transition potentials,
with the probe’s depth of field setting the resolving power.
% with optimal depth precision limited by the probe's depth of field. 
Although dynamical scattering compromises this interpretation, its effects can be substantially mitigated through averaging and off-zone-axis acquisition.

To overcome the scarcity of core-loss events against a momentum-resolved background, we introduce a Bayesian framework that achieves nanometre to sub-nanometre axial and sub-angstrom lateral localization with several hundred core-loss counts in simulation. Bayesian model comparison further demonstrates resolving power beyond the conventional Rayleigh criterion, which is limited by the probe's depth of field.

% \section{Competing interests}
% No competing interest is declared.

\section{Acknowledgments}
The author would like to thank Scott D. Findlay for helpful discussions and suggestions in preparing this work. This research is supported by an Australian Government Research Training Program (RTP) Scholarship.
OpenAI Codex (Sol 6.1) was used for checking code, figure polishing, manuscript proofreading and light edits. The author takes full responsibility for the final manuscript.

\section{Appendix}

\subsection{Derivation of $\hat{I}(\vQ,\vq)$}\label{sec:appendix_IQq_derivation}

Fourier transforming \eqref{eq:I_final} wrt $\vR$
\begin{equation}
\hat{I}(\mathbf{Q},\mathbf{q}) = \int d^2R\; e^{-2\pi i \mathbf{Q}\cdot\mathbf{R}} \sum_{fi} \left|\mathcal{F}\left\{ \mathcal{M}_{z'\rightarrow\text{exit}}\, (-i\sigma_{fi})H_{fi}   \mathcal{M}_{\text{entrance}\rightarrow z'}\,\psi_0(\mathbf{r}-\mathbf{R}) \right\}\right|^2.
\end{equation}
In the Born approximation, the detector-plane wavefunction for channel $fi$ is
\begin{equation}
\begin{split}
    \hat{\psi}_{fi}(\mathbf{R},\mathbf{q};z')
    &= -i\sigma_{fi} \, e^{-i\pi\lambda(t-z')q^2} 
     \\ &\quad\times \int d^2q'\;
     e^{-i2\pi (\vq-\vq')\cdot\vx} \hat{H}_{fi}(\mathbf{q}-\mathbf{q}') \\
    &\quad \times e^{-i\pi\lambda z' q'^2} \,
      \hat{\psi}_0(\mathbf{q}')\,
      e^{-2\pi i\mathbf{q}'\cdot\mathbf{R}} e^{-i\pi\lambda \Delta f q'^2} ,
\end{split}
\label{eq:apexp1}
\end{equation}
For simplicity, in \cref{eq:apexp1} and subsequent expressions, the transmission function will be absorbed into the transition potential $tH\rightarrow H$ and recovered at the end. 
The probe begins at scan position $\mathbf{R}$ corresponding to the linear phase $e^{-2\pi i\mathbf{q}'\cdot\mathbf{R}}$ and defocus $\Delta f$ with negative values corresponding to underfocus.
The Fresnel propagator $e^{-i\pi\lambda z' q'^2}$ advances the probe from the entrance face to the ionization depth $z'$, where the transition operator $H_{fi}$ acts on the propagated probe as in \cref{eq:Hpsi}, generating the inelastic wave via convolution with $\hat{H}_{fi}$ in reciprocal space. 
 The transition potential is shifted laterally to the dopant's position which we denote $\vx$ which manifests as a phase ramp in reciprocal space due to the Fourier shift theorem.
The remaining propagator $e^{-i\pi\lambda(t-z')q^2}$ advances the inelastic wave from $z'$ to the exit face.
It will be instructive to define $z=z'+\Delta f$ since
\begin{align}
 e^{-i\pi\lambda z' q'^2}  e^{-i\pi\lambda \Delta f q'^2} =  e^{-i\pi\lambda (z'+\Delta f) q'^2} 
\end{align}
which is the relative distance of the ionization event from the focal plane.
The leading phase $e^{-i\pi\lambda(t-z')q^2}$ has unit modulus and drops out of $|\hat{\psi}_{fi}|^2$. Expanding the intensity as a product of two such integrals over dummy variables $\mathbf{q}'$ and $\mathbf{q}''$ and substituting into the scan FT,
\begin{equation}
\begin{split}
    \hat{I}(\mathbf{Q},\mathbf{q})
    &=\sum_{fi}  \sigma_{fi}^2
      \iint d^2q'\,d^2q''\;
     e^{-i2\pi (\vq''-\vq')\cdot\vx} \\ 
     &\quad \times \hat{H}_{fi}(\mathbf{q}-\mathbf{q}')\,\hat{H}_{fi}^*(\mathbf{q}-\mathbf{q}'')\, \\
    &\quad \times\hat{\psi}_0(\mathbf{q}')\,\hat{\psi}_0^*(\mathbf{q}'')\,
      e^{-i\pi\lambda z(q'^2-q''^2)}
      \\ &\quad \times \int d^2R\;e^{-2\pi i(\mathbf{Q}+\mathbf{q}'-\mathbf{q}'')\cdot\mathbf{R}}.
\end{split}
\end{equation}
The $R$-integral is a Fourier representation of the delta function,
$\delta^{(2)}(\mathbf{q}''-\mathbf{q}'-\mathbf{Q})$, which collapses the $\mathbf{q}''$ integral
by setting $\mathbf{q}''=\mathbf{q}'+\mathbf{Q}$. The Fresnel phase difference then simplifies as
\begin{equation}
    q''^2 - q'^2
    = |\mathbf{q}'+\mathbf{Q}|^2 - q'^2
    = 2\mathbf{q}'\cdot\mathbf{Q} + Q^2,
\end{equation}
yielding
\begin{equation}
\begin{split}
    \hat{I}(\mathbf{Q},\mathbf{q})
    &= \sum_{fi} \sigma_{fi}^2 e^{-i2\pi \vQ\cdot\vx} 
      \\&\quad \times\int d^2q'\;  
      \hat{H}_{fi}(\mathbf{q}-\mathbf{q}')\,\hat{H}_{fi}^*(\mathbf{q}-\mathbf{q}'-\mathbf{Q}) \\
    &\quad \times \hat{\psi}_0(\mathbf{q}')\,\hat{\psi}_0^*(\mathbf{q}'+\mathbf{Q})\;
      e^{i\pi\lambda z(2\mathbf{q}'\cdot\mathbf{Q}+Q^2)}.
\end{split}
\end{equation}
Substituting $\mathbf{u} = \mathbf{q} - \mathbf{q}'$ (the momentum transferred through $H_{fi}$),
so that $\mathbf{q}' = \mathbf{q} - \mathbf{u}$, the Fresnel phase exponent becomes
\begin{equation}
    2\mathbf{q}'\cdot\mathbf{Q} + Q^2
    = 2(\mathbf{q}-\mathbf{u})\cdot\mathbf{Q} + Q^2
    = (2\mathbf{q}\cdot\mathbf{Q} + Q^2) - 2\mathbf{u}\cdot\mathbf{Q},
\end{equation}
separating into a detector-coordinate prefactor and a $\mathbf{u}$-dependent phase inside the integral. 
Collecting all factors,
\begin{align}
    \hat{I}(\mathbf{Q},\mathbf{q})
    &=  e^{-2\pi i\mathbf{Q}\cdot\mathbf{x}}\,
       e^{\,i\pi\lambda z(2\mathbf{q}\cdot\mathbf{Q}+Q^2)} \nonumber\\
    &\quad\times \int d^2u\;
      \sum_{fi} \sigma_{fi}^2 \hat{H}_{fi}(\mathbf{u})\,\hat{H}_{fi}^*(\mathbf{u}-\mathbf{Q})\,
      e^{-2\pi i\lambda z\mathbf{u}\cdot\mathbf{Q}} \nonumber\\
    &\qquad\times
      \hat{\psi}_0(\mathbf{q}-\mathbf{u})\,\hat{\psi}_0^*(\mathbf{q}-\mathbf{u}+\mathbf{Q}).
\end{align}
Applying a further shift $\mathbf{u} \rightarrow \mathbf{u} + \mathbf{Q}/2$ centres the transition potential arguments symmetrically, giving
\begin{align}
    \hat{I}(\mathbf{Q},\mathbf{q})
    &= e^{-2\pi i\mathbf{Q}\cdot\mathbf{x}}\,
       e^{\,2i\pi\lambda z\mathbf{q}\cdot\mathbf{Q}} \nonumber
     \int d^2u\; e^{-2\pi i\lambda z\mathbf{u}\cdot\mathbf{Q}} \\ &\quad\times  
      \sum_{fi} \sigma_{fi}^2  (\hat t \conv \hat{H}_{fi})\!\left(\mathbf{u}+\tfrac{\mathbf{Q}}{2}\right)
     (\hat t \conv \hat{H}_{fi})^*\!\left(\mathbf{u}-\tfrac{\mathbf{Q}}{2}\right)
       \nonumber\\
    &\qquad\times
      \hat{\psi}_0\!\left(\mathbf{q}-\mathbf{u}-\tfrac{\mathbf{Q}}{2}\right)
      \hat{\psi}_0^*\!\left(\mathbf{q}-\mathbf{u}+\tfrac{\mathbf{Q}}{2}\right).
\end{align}
with the transmission function reinstated.

\subsection{Rotational invariance of $\hat {\mathcal{H}}(\vQ,\vxi)$}\label{sec:Hn0_iso_proof}

We prove the rotational invariance of the transition potential ambiguity functions using the explicit representation of \cite{dwyer2005multislice}.
In the central-potential model, the initial bound state $|nlm\rangle$ is parametrized by a principal quantum number, orbital and magnetic quantum numbers, respectively.
The final continuum state $|\varepsilon_f l'm'\rangle$ is parametrized by a continuous analogue of the principal quantum number $\varepsilon_f$  corresponding to the energy above the ionization threshold imparted, and an orbital and magnetic quantum numbers respectively.

Following \cite{dwyer2005multislice}, we decompose the bound state and continuum wavefunctions into a radial and an angular component
\begin{align}
\varphi_{nlm}(\vx)&=\frac{P_{nl}(|\vx|)}{|\vx|}\,Y_l^{m}(\hat\vx),
\\
\varphi_{\varepsilon_f l'm'}(\vx)&=\frac{P_{\varepsilon_f l'}(|\vx|)}{|\vx|}\,Y_{l'}^{m'}(\hat\vx).
\label{eq:radangwf}
\end{align}
where $Y_l^m$ denotes a spherical harmonic.
With this representation and set of assumptions, in our Fourier convention, \cite{dwyer2005multislice} obtains
\begin{equation}
\hat H_{fi}(\vq)=\frac{4e^2}{|\vq|^2}\,\eta_{fi}(\vq)
\label{eq:etaVfi}
\end{equation}
where
\begin{equation}
\eta_{fi}(\vq)=\sum_{l''=0}^{\infty}\,\sum_{m''=-l''}^{+l''}\,(-i)^{l''}\,\begin{bmatrix} l''\,m'' \\ l'\,m',\,l\,m \end{bmatrix}\,Y_{l''}^{m''}(\hat\vq)\,\mathcal I_{l''}(|\vq|)
\label{eq:eta13}
\end{equation}
where $\mathcal I_{l''}(|\vq|)=\int_0^\infty dr\,P_{\varepsilon_f l'}(r)\,j_{l''}(2\pi|\vq|r)\,P_{nl}(r)$ which is a function of $|\vq|$ only. So
\begin{align}
\hat{\mathcal H}_{nl\to\varepsilon_f l'}(\vQ,\vv)
&=\sum_{m=-l}^{l}\sum_{m'=-l'}^{l'}\hat H_{fi}(\vq_+)\,\hat H_{fi}^{*}(\vq_-)
\nonumber\\
&=\frac{16e^4}{|\vq_+|^2|\vq_-|^2}
\sum_{m,m'}\eta_{fi}(\vq_+)\,\eta_{fi}^{*}(\vq_-)  \nonumber\\
&=\frac{16e^4}{|\vq_+|^2|\vq_-|^2}
\sum_{l'',L}\sum_{m'',M}(-i)^{l''}\,i^{L}
\nonumber\\
&\qquad\times
\mathcal I_{l''}(|\vq_+|)\,\mathcal I_{L}(|\vq_-|)
\nonumber\\
&\qquad\times
Y_{l''}^{m''}(\hat\vq_+)\,Y_{L}^{M*}(\hat\vq_-)
\nonumber\\
&\qquad\times
\sum_{m,m'}
\begin{bmatrix} l''m'' \\ l'm',lm \end{bmatrix}
\begin{bmatrix} LM \\ l'm',lm \end{bmatrix}
\label{eq:HHstar_mid}
\end{align}
where $\vq_\pm=(\vv\pm\tfrac{\vQ}{2},\,q_z)$ are three-dimensional vectors, with $\vv,\vQ$ the in-plane components. The rotation $R$ acts only on the components perpendicular to the optic axis, $(\vq_\perp,q_z)\to(R\vq_\perp,q_z)$, since $q_z=\Delta E/hv$ is fixed by energy conservation where $\Delta E$ is the total energy lost by the beam electron to the inelastic event. Hence $q_\pm$ and $\hat\vq_+\!\cdot\hat\vq_-$ are unchanged by $R$.
% Product of the two angular integrals, written out
\begin{equation}
\begin{aligned}
\sum_{m,m'} \begin{bmatrix} l''m'' \\ l'm',lm \end{bmatrix} \begin{bmatrix} LM \\ l'm',lm \end{bmatrix} &=(-1)^{m''+M}\,\frac{(2l'+1)(2l+1)}{4\pi} \\
&\quad \times \sqrt{(2l''+1)(2L+1)}\, \begin{pmatrix} l' & l'' & l\\ 0&0&0\end{pmatrix} \begin{pmatrix} l' & L & l\\ 0&0&0\end{pmatrix} \\
&\quad \times\sum_{m,m'} \begin{pmatrix} l' & l'' & l\\ -m'&-m''&m\end{pmatrix} \begin{pmatrix} l' & L & l\\ -m'&-M&m\end{pmatrix},
\end{aligned}
\label{eq:sub_step1}
\end{equation}
where the phase $(-1)^{2m'}=1$ because $m'$ is an integer.
% Reorder columns so the summed slots are first and second
Cyclically or pairwise permuting the columns of a $3j$ symbol changes it by at most the sign $(-1)^{l'+l''+l}$. Swapping the second and third columns of both symbols gives
\begin{equation}
\begin{pmatrix} l' & l'' & l\\ -m'&-m''&m\end{pmatrix}
=(-1)^{l'+l''+l}
\begin{pmatrix} l' & l & l''\\ -m'&m&-m''\end{pmatrix},
\end{equation}
and likewise with $(l'',m'')\to(L,M)$. The two sign factors multiply to $(-1)^{2l'+2l+l''+L}$, which equals $1$ once $L=l''$ is enforced (and the $3j$ symbols with all-zero bottom row vanish unless $l'+l+l''$ is even anyway).
% Apply orthogonality: summed slots are (-m', m), free slot is the l'' column
Now the summed slots are the first two columns, with $(m_1,m_2)=(-m',m)$ and free label $m_3=-m''$. The orthogonality relation is given by
\begin{equation}
\sum_{m,m'}
\begin{pmatrix} l' & l & l''\\ -m'&m&-m''\end{pmatrix}
\begin{pmatrix} l' & l & L\\ -m'&m&-M\end{pmatrix}
=\frac{\delta_{l''L}\,\delta_{m''M}}{2l''+1}.
\label{eq:sub_ortho}
\end{equation}
% Combine
Inserting \eqref{eq:sub_ortho} into \eqref{eq:sub_step1}, the phase is $(-1)^{m''+M}=(-1)^{2m''}=1$ and $\sqrt{(2l''+1)(2L+1)}/(2l''+1)=1$, so
\begin{equation}
\sum_{m,m'} \begin{bmatrix} l''m'' \\ l'm',lm \end{bmatrix} \begin{bmatrix} LM \\ l'm',lm \end{bmatrix} =\delta_{l''L}\,\delta_{m''M}\, \times\frac{(2l'+1)(2l+1)}{4\pi} \begin{pmatrix} l' & l'' & l\\ 0&0&0\end{pmatrix}^{2}.
\label{eq:sublevel_new}
\end{equation}
Inserting \cref{eq:sublevel_new} back into \cref{eq:HHstar_mid}, the Kronecker deltas set
$L=l''$ and $M=m''$. The phases combine as $(-i)^{l''}\,i^{l''}=1$, and since the radial functions and spherical Bessel functions are real, $\mathcal I_{l''}$ is real. Therefore 
\begin{align}
\hat{\mathcal H}_{nl\to\varepsilon_f l'}(\vQ,\vv)
&=\frac{16e^4}{|\vq_+|^2|\vq_-|^2}\,
\frac{(2l'+1)(2l+1)}{4\pi}
\nonumber\\
&\quad\times\sum_{l''}
\begin{pmatrix} l' & l'' & l\\ 0&0&0\end{pmatrix}^{2}
\mathcal I_{l''}(|\vq_+|)\,\mathcal I_{l''}(|\vq_-|)
\nonumber\\
&\quad\times\sum_{m''=-l''}^{l''}Y_{l''}^{m''}(\hat\vq_+)\,Y_{l''}^{m''*}(\hat\vq_-).
\label{eq:HHstar_collapsed_new}
\end{align}
The only remaining dependence on the directions $\hat\vq_\pm$ is through the $m''$ sum, which the spherical-harmonic addition theorem
\begin{equation}
\sum_{m''=-l''}^{l''}Y_{l''}^{m''}(\hat\vq_+)\,Y_{l''}^{m''*}(\hat\vq_-)
=\frac{2l''+1}{4\pi}\,P_{l''}(\hat\vq_+\!\cdot\hat\vq_-)
\end{equation}
reduces to a Legendre polynomial of the angle between $\vq_+$ and $\vq_-$. Here $P_{l''}$ is a Legendre polynomial, not to be confused with the radial wavefunctions $P_{nl}$ and $P_{\varepsilon_f l'}$ which are distinguished by their subscripts.

Consequently the transition potentials ambiguity function depends on $\vq_+,\vq_-$ only through $q_+$, $q_-$ and $\hat\vq_+\!\cdot\hat\vq_-$. Hence, given some rotation $R\in {\rm SO}(2)$ 
% \begin{align}
%     \hat{\mathcal H}(R\vQ,R\vv)=\hat{\mathcal H}(\vQ,\vv).
% \end{align}
% Now
\begin{equation}
\hat{\mathcal H}(R\vQ,R\vxi)=\int d^2v\;e^{-2\pi i\,\vv\cdot R\vxi}\,\hat{\mathcal H}(R\vQ,\vv),
\end{equation}
changing variables to $\vv=R\vv'$, with $|\det R|=1$ and $R\vv'\cdot R\vxi=\vv'\cdot\vxi$ gives
\begin{align}
\hat{\mathcal H}(R\vQ,R\vxi)
&=\int d^2v'\;e^{-2\pi i\,\vv'\cdot\vxi}\,\hat{\mathcal H}(R\vQ,R\vv')
\nonumber\\
&=\int d^2v'\;e^{-2\pi i\,\vv'\cdot\vxi}\,\hat{\mathcal H}(\vQ,\vv')
\nonumber\\
&=\hat{\mathcal H}(\vQ,\vxi).
\end{align}
Hence $\hat{\mathcal H}$ depends only on $Q$, $\xi$ and the angle between them. This result was proven for only a single transition, but generalizes trivially.

\clearpage
\bibliographystyle{abbrvnat}

\bibliography{references}

\begin{thebibliography}{46}
\providecommand{\natexlab}[1]{#1}
\providecommand{\url}[1]{\texttt{#1}}
\expandafter\ifx\csname urlstyle\endcsname\relax
  \providecommand{\doi}[1]{doi: #1}\else
  \providecommand{\doi}{doi: \begingroup \urlstyle{rm}\Url}\fi

\bibitem[Arslan et~al.(2006)Arslan, Tong, and Midgley]{arslan2006reducing}
I.~Arslan, J.~R. Tong, and P.~A. Midgley.
\newblock Reducing the missing wedge: High-resolution dual axis tomography of
  inorganic materials.
\newblock \emph{Ultramicroscopy}, 106\penalty0 (11-12):\penalty0 994--1000,
  2006.

\bibitem[Borisevich et~al.(2006)Borisevich, Lupini, and
  Pennycook]{borisevich2006depth}
A.~Y. Borisevich, A.~R. Lupini, and S.~J. Pennycook.
\newblock Depth sectioning with the aberration-corrected scanning transmission
  electron microscope.
\newblock \emph{Proceedings of the National Academy of Sciences}, 103\penalty0
  (9):\penalty0 3044--3048, 2006.

\bibitem[Born and Wolf(1999)]{born1999principles}
M.~Born and E.~Wolf.
\newblock \emph{Principles of Optics: Electromagnetic Theory of Propagation,
  Interference and Diffraction of Light}.
\newblock Cambridge University Press, 7th edition, 1999.

\bibitem[Brown et~al.(2018)Brown, Chen, Weyland, Ophus, Ciston, Allen, and
  Findlay]{brown2018structure}
H.~G. Brown, Z.~Chen, M.~Weyland, C.~Ophus, J.~Ciston, L.~J. Allen, and S.~D.
  Findlay.
\newblock Structure retrieval at atomic resolution in the presence of multiple
  scattering of the electron probe.
\newblock \emph{Physical Review Letters}, 121\penalty0 (26):\penalty0 266102,
  2018.

\bibitem[Deimetry et~al.(2025)Deimetry, Petersen, Weyland, and
  Findlay]{deimetry2025dopant}
M.~Deimetry, T.~C. Petersen, M.~Weyland, and S.~D. Findlay.
\newblock Dopant site occupancy determined by core-loss-filtered,
  position-averaged convergent beam electron diffraction.
\newblock \emph{Microscopy and Microanalysis}, 31\penalty0 (6):\penalty0
  ozaf124, 2025.

\bibitem[Deimetry et~al.(2026{\natexlab{a}})Deimetry, Petersen, Weyland, and
  Findlay]{deimetry2026threedimensionaldopantdistributionsingle}
M.~Deimetry, T.~Petersen, M.~Weyland, and S.~Findlay.
\newblock Three-dimensional dopant distribution from a single, out-of-focus
  spectroscopic image, 2026{\natexlab{a}}.
\newblock arXiv:2610.11053.

\bibitem[Deimetry et~al.(2026{\natexlab{b}})Deimetry, Petersen, Weyland, and
  Findlay]{deimetry2026bayesian}
M.~Deimetry, T.~C. Petersen, M.~Weyland, and S.~D. Findlay.
\newblock A {Bayesian} approach to spectroscopic depth sectioning for locating
  dopant atoms.
\newblock \emph{Journal of Microscopy}, 2026{\natexlab{b}}.

\bibitem[Dwyer(2005)]{dwyer2005multislice}
C.~Dwyer.
\newblock Multislice theory of fast electron scattering incorporating atomic
  inner-shell ionization.
\newblock \emph{Ultramicroscopy}, 104\penalty0 (2):\penalty0 141--151, 2005.

\bibitem[Dwyer(2015)]{DWYER201568}
C.~Dwyer.
\newblock The role of symmetry in the theory of inelastic high-energy electron
  scattering and its application to atomic-resolution core-loss imaging.
\newblock \emph{Ultramicroscopy}, 151:\penalty0 68--77, 2015.
\newblock ISSN 0304-3991.
\newblock \doi{https://doi.org/10.1016/j.ultramic.2014.11.007}.
\newblock Special Issue: 80th Birthday of Harald Rose; PICO 2015 – Third
  Conference on Frontiers of Aberration Corrected Electron Microscopy.

\bibitem[D’Alfonso et~al.(2007)D’Alfonso, Findlay, Oxley, Pennycook, van
  Benthem, and Allen]{d2007depth}
A.~J. D’Alfonso, S.~D. Findlay, M.~P. Oxley, S.~J. Pennycook, K.~van Benthem,
  and L.~J. Allen.
\newblock Depth sectioning in scanning transmission electron microscopy based
  on core-loss spectroscopy.
\newblock \emph{Ultramicroscopy}, 108\penalty0 (1):\penalty0 17--28, 2007.

\bibitem[Egerton(1975)]{egerton1975inelastic}
R.~F. Egerton.
\newblock Inelastic scattering of 80 {keV} electrons in amorphous carbon.
\newblock \emph{Philosophical Magazine}, 31\penalty0 (1):\penalty0 199--215,
  1975.

\bibitem[Egerton and Malac(2002)]{egerton2002improved}
R.~F. Egerton and M.~Malac.
\newblock Improved background-fitting algorithms for ionization edges in
  electron energy-loss spectra.
\newblock \emph{Ultramicroscopy}, 92\penalty0 (2):\penalty0 47--56, 2002.

\bibitem[Ercius et~al.(2024)Ercius, Johnson, Pelz, Savitzky, Hughes, Brown,
  Zeltmann, Hsu, Pedroso, Cohen, et~al.]{ercius20244d}
P.~Ercius, I.~J. Johnson, P.~Pelz, B.~H. Savitzky, L.~Hughes, H.~G. Brown,
  S.~E. Zeltmann, S.-L. Hsu, C.~C.~S. Pedroso, B.~E. Cohen, et~al.
\newblock The {4D} camera: An 87 {kHz} direct electron detector for
  scanning/transmission electron microscopy.
\newblock \emph{Microscopy and {Microanalysis}}, 30\penalty0 (5):\penalty0
  903--912, 2024.

\bibitem[Gao et~al.(2017)Gao, Wang, Zhang, Martinez, Nellist, Pan, and
  Kirkland]{gao2017electron}
S.~Gao, P.~Wang, F.~Zhang, G.~T. Martinez, P.~D. Nellist, X.~Pan, and A.~I.
  Kirkland.
\newblock Electron ptychographic microscopy for three-dimensional imaging.
\newblock \emph{Nature {Communications}}, 8\penalty0 (1):\penalty0 163, 2017.

\bibitem[Jeffreys(1998)]{jeffreys1998theory}
H.~Jeffreys.
\newblock \emph{The theory of probability}.
\newblock Oxford University Press, 1998.

\bibitem[Kass and Raftery(1995)]{kass1995bayes}
R.~E. Kass and A.~E. Raftery.
\newblock Bayes factors.
\newblock \emph{Journal of the american statistical association}, 90\penalty0
  (430):\penalty0 773--795, 1995.

\bibitem[Koposov et~al.(2026)Koposov, Speagle, Barbary, Ashton, Bennett,
  Buchner, Scheffler, Talbot, Cook, Guillochon, Cubillos, Ramos, Dartiailh,
  Ilya, Tollerud, Lang, Johnson, jtmendel, Higson, Vandal, Daylan, Angus,
  ruskin23, Cargile, Sheehan, Pitkin, Kirk, Xu, Leja, and
  joezuntz]{sergey_koposov_2026_21417233}
S.~Koposov, J.~Speagle, K.~Barbary, G.~Ashton, E.~Bennett, J.~Buchner,
  C.~Scheffler, C.~Talbot, B.~Cook, J.~Guillochon, P.~Cubillos, A.~A. Ramos,
  M.~Dartiailh, Ilya, E.~Tollerud, D.~Lang, B.~Johnson, jtmendel, E.~Higson,
  T.~Vandal, T.~Daylan, R.~Angus, ruskin23, P.~Cargile, P.~Sheehan, M.~Pitkin,
  M.~Kirk, L.~Xu, J.~Leja, and joezuntz.
\newblock joshspeagle/dynesty: v3.1.0, July 2026.

\bibitem[Leary et~al.(2013)Leary, Saghi, Midgley, and
  Holland]{leary2013compressed}
R.~Leary, Z.~Saghi, P.~A. Midgley, and D.~J. Holland.
\newblock Compressed sensing electron tomography.
\newblock \emph{Ultramicroscopy}, 131:\penalty0 70--91, 2013.

\bibitem[Lin et~al.(2015)Lin, Teng, Chiu, and Suenaga]{lin2015exploring}
Y.-C. Lin, P.-Y. Teng, P.-W. Chiu, and K.~Suenaga.
\newblock Exploring the single atom spin state by electron spectroscopy.
\newblock \emph{Physical Review Letters}, 115\penalty0 (20):\penalty0 206803,
  2015.

\bibitem[Liu and Brown(1987)]{liu1987influence}
D.-R. Liu and L.~Brown.
\newblock Influence of some practical factors on background extrapolation in
  {EELS} quantification.
\newblock \emph{Journal of Microscopy}, 147\penalty0 (1):\penalty0 37--49,
  1987.

\bibitem[Lupini and De~Jonge(2011)]{lupini2011three}
A.~R. Lupini and N.~De~Jonge.
\newblock The three-dimensional point spread function of aberration-corrected
  scanning transmission electron microscopy.
\newblock \emph{Microscopy and Microanalysis}, 17\penalty0 (5):\penalty0
  817--826, 2011.

\bibitem[Maiden and Rodenburg(2009)]{maiden2009improved}
A.~M. Maiden and J.~M. Rodenburg.
\newblock An improved ptychographical phase retrieval algorithm for diffractive
  imaging.
\newblock \emph{Ultramicroscopy}, 109\penalty0 (10):\penalty0 1256--1262, 2009.

\bibitem[Maiden et~al.(2012)Maiden, Humphry, and
  Rodenburg]{maiden2012ptychographic}
A.~M. Maiden, M.~J. Humphry, and J.~M. Rodenburg.
\newblock Ptychographic transmission microscopy in three dimensions using a
  multi-slice approach.
\newblock \emph{Journal of the Optical Society of America {A}}, 29\penalty0
  (8):\penalty0 1606--1614, 2012.

\bibitem[Midgley et~al.(1995)Midgley, Saunders, Vincent, and
  Steeds]{midgley1995energy}
P.~A. Midgley, M.~Saunders, R.~Vincent, and J.~W. Steeds.
\newblock Energy-filtered convergent-beam diffraction: examples and future
  prospects.
\newblock \emph{Ultramicroscopy}, 59:\penalty0 1--13, 1995.

\bibitem[M{\"u}ller-Caspary et~al.(2017)M{\"u}ller-Caspary, Krause, Grieb,
  L{\"o}ffler, Schowalter, B{\'e}ch{\'e}, Galioit, Marquardt, Zweck,
  Schattschneider, et~al.]{muller2017measurement}
K.~M{\"u}ller-Caspary, F.~F. Krause, T.~Grieb, S.~L{\"o}ffler, M.~Schowalter,
  A.~B{\'e}ch{\'e}, V.~Galioit, D.~Marquardt, J.~Zweck, P.~Schattschneider,
  et~al.
\newblock Measurement of atomic electric fields and charge densities from
  average momentum transfers using scanning transmission electron microscopy.
\newblock \emph{Ultramicroscopy}, 178:\penalty0 62--80, 2017.

\bibitem[Ophus(2019)]{ophus2019four}
C.~Ophus.
\newblock Four-dimensional scanning transmission electron microscopy
  ({4D}-{STEM}): From scanning nanodiffraction to ptychography and beyond.
\newblock \emph{Microscopy and {Microanalysis}}, 25\penalty0 (3):\penalty0
  563--582, 2019.

\bibitem[Pelz et~al.(2021)Pelz, Brown, Stonemeyer, Findlay, Zettl, Ercius,
  Zhang, Ciston, Scott, and Ophus]{pelz2021phase}
P.~M. Pelz, H.~G. Brown, S.~Stonemeyer, S.~D. Findlay, A.~Zettl, P.~Ercius,
  Y.~Zhang, J.~Ciston, M.~C. Scott, and C.~Ophus.
\newblock Phase-contrast imaging of multiply-scattering extended objects at
  atomic resolution by reconstruction of the scattering matrix.
\newblock \emph{Physical Review Research}, 3\penalty0 (2):\penalty0 023159,
  2021.

\bibitem[Pennycook et~al.(2015)Pennycook, Lupini, Yang, Murfitt, Jones, and
  Nellist]{pennycook2015efficient}
T.~J. Pennycook, A.~R. Lupini, H.~Yang, M.~F. Murfitt, L.~Jones, and P.~D.
  Nellist.
\newblock Efficient phase contrast imaging in {STEM} using a pixelated
  detector. {Part} 1: Experimental demonstration at atomic resolution.
\newblock \emph{Ultramicroscopy}, 151:\penalty0 160--167, 2015.

\bibitem[Philipp et~al.(2022)Philipp, Tate, Shanks, Mele, Peemen, Dona,
  Hartong, van Veen, Shao, Chen, et~al.]{philipp2022very}
H.~T. Philipp, M.~W. Tate, K.~S. Shanks, L.~Mele, M.~Peemen, P.~Dona,
  R.~Hartong, G.~van Veen, Y.-T. Shao, Z.~Chen, et~al.
\newblock Very-high dynamic range, 10,000 frames/second pixel array detector
  for electron microscopy.
\newblock \emph{Microscopy and Microanalysis}, 28\penalty0 (2):\penalty0
  425--440, 2022.

\bibitem[Pun et~al.(1984)Pun, Ellis, and Eden]{pun1984optimized}
T.~Pun, J.~R. Ellis, and M.~Eden.
\newblock Optimized acquisition parameters and statistical detection limit in
  quantitative {EELS}.
\newblock \emph{Journal of {Microscopy}}, 135\penalty0 (3):\penalty0 295--316,
  1984.

\bibitem[Pun et~al.(1985)Pun, Ellis, and Eden]{pun1985weighted}
T.~Pun, J.~R. Ellis, and M.~Eden.
\newblock Weighted least squares estimation of background in {EELS} imaging.
\newblock \emph{Journal of {Microscopy}}, 137\penalty0 (1):\penalty0 93--100,
  1985.

\bibitem[Ramasse et~al.(2013)Ramasse, Seabourne, Kepaptsoglou, Zan, Bangert,
  and Scott]{ramasse2013probing}
Q.~M. Ramasse, C.~R. Seabourne, D.-M. Kepaptsoglou, R.~Zan, U.~Bangert, and
  A.~J. Scott.
\newblock Probing the bonding and electronic structure of single atom dopants
  in graphene with electron energy loss spectroscopy.
\newblock \emph{Nano {Letters}}, 13\penalty0 (10):\penalty0 4989--4995, 2013.

\bibitem[Rodenburg and Bates(1992)]{rodenburg1992theory}
J.~M. Rodenburg and R.~H.~T. Bates.
\newblock The theory of super-resolution electron microscopy via
  {Wigner}-distribution deconvolution.
\newblock \emph{Philosophical Transactions: Physical Sciences and Engineering},
  pages 521--553, 1992.

\bibitem[Rodenburg and Faulkner(2004)]{rodenburg2004phase}
J.~M. Rodenburg and H.~M.~L. Faulkner.
\newblock A phase retrieval algorithm for shifting illumination.
\newblock \emph{Applied {Physics} {Letters}}, 85\penalty0 (20):\penalty0
  4795--4797, 2004.

\bibitem[Saghi et~al.(2011)Saghi, Holland, Leary, Falqui, Bertoni, Sederman,
  Gladden, and Midgley]{saghi2011three}
Z.~Saghi, D.~J. Holland, R.~Leary, A.~Falqui, G.~Bertoni, A.~J. Sederman, L.~F.
  Gladden, and P.~A. Midgley.
\newblock Three-dimensional morphology of iron oxide nanoparticles with
  reactive concave surfaces. a compressed sensing-electron tomography ({CS-ET})
  approach.
\newblock \emph{Nano Letters}, 11\penalty0 (11):\penalty0 4666--4673, 2011.

\bibitem[Scott et~al.(2012)Scott, Chen, Mecklenburg, Zhu, Xu, Ercius, Dahmen,
  Regan, and Miao]{scott2012electron}
M.~C. Scott, C.-C. Chen, M.~Mecklenburg, C.~Zhu, R.~Xu, P.~Ercius, U.~Dahmen,
  B.~C. Regan, and J.~Miao.
\newblock Electron tomography at 2.4-{\aa}ngstr{\"o}m resolution.
\newblock \emph{Nature}, 483\penalty0 (7390):\penalty0 444--447, 2012.

\bibitem[{Speagle}(2020)]{2020MNRAS.493.3132S}
J.~S. {Speagle}.
\newblock {DYNESTY: a dynamic nested sampling package for estimating Bayesian
  posteriors and evidences}.
\newblock \emph{\mnras}, 493\penalty0 (3):\penalty0 3132--3158, Apr. 2020.
\newblock \doi{10.1093/mnras/staa278}.

\bibitem[Su and Zeitler(1993)]{su1993background}
D.~S. Su and E.~Zeitler.
\newblock Background problem in electron-energy-loss spectroscopy.
\newblock \emph{Physical Review B}, 47\penalty0 (22):\penalty0 14734, 1993.

\bibitem[Suenaga et~al.(2000)Suenaga, Tence, Mory, Colliex, Kato, Okazaki,
  Shinohara, Hirahara, Bandow, and Iijima]{suenaga2000element}
K.~Suenaga, M.~Tence, C.~Mory, C.~Colliex, H.~Kato, T.~Okazaki, H.~Shinohara,
  K.~Hirahara, S.~Bandow, and S.~Iijima.
\newblock Element-selective single atom imaging.
\newblock \emph{Science}, 290\penalty0 (5500):\penalty0 2280--2282, 2000.
\newblock \doi{https://doi.org/10.1126/science.290.5500.2280}.

\bibitem[Suenaga et~al.(2009)Suenaga, Sato, Liu, Kataura, Okazaki, Kimoto,
  Sawada, Sasaki, Omoto, Tomita, et~al.]{suenaga2009visualizing}
K.~Suenaga, Y.~Sato, Z.~Liu, H.~Kataura, T.~Okazaki, K.~Kimoto, H.~Sawada,
  T.~Sasaki, K.~Omoto, T.~Tomita, et~al.
\newblock Visualizing and identifying single atoms using electron energy-loss
  spectroscopy with low accelerating voltage.
\newblock \emph{Nature {Chemistry}}, 1\penalty0 (5):\penalty0 415--418, 2009.

\bibitem[Tate et~al.(2016)Tate, Purohit, Chamberlain, Nguyen, Hovden, Chang,
  Deb, Turgut, Heron, Schlom, et~al.]{tate2016high}
M.~W. Tate, P.~Purohit, D.~Chamberlain, K.~X. Nguyen, R.~Hovden, C.~S. Chang,
  P.~Deb, E.~Turgut, J.~T. Heron, D.~G. Schlom, et~al.
\newblock High dynamic range pixel array detector for scanning transmission
  electron microscopy.
\newblock \emph{Microscopy and Microanalysis}, 22\penalty0 (1):\penalty0
  237--249, 2016.

\bibitem[Terzoudis-Lumsden et~al.(2023)Terzoudis-Lumsden, Petersen, Brown,
  Pelz, Ophus, and Findlay]{terzoudis2023resolution}
E.~W.~C. Terzoudis-Lumsden, T.~C. Petersen, H.~G. Brown, P.~M. Pelz, C.~Ophus,
  and S.~D. Findlay.
\newblock Resolution of virtual depth sectioning from four-dimensional scanning
  transmission electron microscopy.
\newblock \emph{Microscopy and Microanalysis}, 29\penalty0 (4):\penalty0
  1409--1421, 2023.

\bibitem[Unser et~al.(1987)Unser, Ellis, Pun, and Eden]{unser1987optimal}
M.~Unser, J.~R. Ellis, T.~Pun, and M.~Eden.
\newblock Optimal background estimation in {EELS}.
\newblock \emph{Journal of {Microscopy}}, 145\penalty0 (Pt 3):\penalty0
  245--256, 1987.

\bibitem[van Benthem et~al.(2006)van Benthem, Lupini, Oxley, Findlay, Allen,
  and Pennycook]{van2006three}
K.~van Benthem, A.~R. Lupini, M.~P. Oxley, S.~D. Findlay, L.~J. Allen, and
  S.~J. Pennycook.
\newblock Three-dimensional {ADF} imaging of individual atoms by through-focal
  series scanning transmission electron microscopy.
\newblock \emph{Ultramicroscopy}, 106\penalty0 (11-12):\penalty0 1062--1068,
  2006.

\bibitem[Verbeeck and Van~Aert(2004)]{verbeeck2004model}
J.~Verbeeck and S.~Van~Aert.
\newblock Model based quantification of {EELS} spectra.
\newblock \emph{Ultramicroscopy}, 101\penalty0 (2-4):\penalty0 207--224, 2004.

\bibitem[Weyland and Midgley(2004)]{weyland2004electron}
M.~Weyland and P.~A. Midgley.
\newblock Electron tomography.
\newblock \emph{Materials Today}, 7\penalty0 (12):\penalty0 32--40, 2004.

\end{thebibliography}

\end{document}